\documentclass[letterpaper, 11pt]{article} 
\usepackage[top=1.0in, bottom=1.0in, left=1.0in, right=1.0in]{geometry}
\usepackage{amsmath, amsfonts, amsthm, amssymb}
\usepackage{url}

\newcommand{\E}{\mathbb{E}}

\newcommand{\R}{\mathbb{R}}

\newcommand{\N}{\mathbb{N}}
\newcommand{\C}{\mathcal{C}}

\renewcommand{\P}{\mathcal{P}}

\newcommand{\eps}{\epsilon}

\renewcommand\vec[1]{\ensuremath\boldsymbol{#1}}

\newcommand\pref[1]{#1}

\renewcommand{\paragraph}[1]{\medskip\noindent{\bf #1}}

\makeatletter
\newtheorem*{rep@theorem}{\rep@title}
\newcommand{\newreptheorem}[2]{%
\newenvironment{rep#1}[1]{%
  \def\rep@title{#2 \ref{##1}}%
  \begin{rep@theorem}}%
  {\end{rep@theorem}}}
\makeatother

\newtheorem{theorem}{Theorem}

\newtheorem{corollary}[theorem]{Corollary}
\newtheorem{lemma}[theorem]{Lemma}
\newtheorem{claim}[theorem]{Claim}

\newtheorem{definition}[theorem]{Definition}

\newreptheorem{theorem}{Theorem}
\newreptheorem{proposition}{Proposition}
\newreptheorem{corollary}{Corollary}
\newreptheorem{lemma}{Lemma}
\newreptheorem{claim}{Claim}
\newreptheorem{conjecture}{Conjecture}

\theoremstyle{definition}

\newtheorem{example}[theorem]{Example}

\begin{document}

\begin{titlepage}

\title{Voting with Coarse Beliefs}
\setcounter{page}{0}

\author{Samantha Leung, Edward Lui, and Rafael Pass \\
Department of Computer Science, Cornell University \\
\texttt{\{samlyy,luied,rafael\}@cs.cornell.edu}}

\maketitle

\begin{abstract}
The classic Gibbard-Satterthwaite theorem says that every strategy-proof voting rule with at least three possible candidates must be dictatorial. Similar impossibility results hold even if we consider a weaker notion of strategy-proofness where voters believe that the other voters' preferences are i.i.d.~(independent and identically distributed). In this paper, we take a bounded-rationality approach to this problem and consider a setting where voters have ``coarse'' beliefs (a notion that has gained popularity in the behavioral economics literature). In particular, we construct good voting rules that satisfy a notion of strategy-proofness with respect to coarse i.i.d.~beliefs, thus circumventing the above impossibility results.
\end{abstract}
\end{titlepage}

\section{Introduction}

People have long desired to have a good voting rule that is \emph{strategy-proof}---that is, the voters would not want to lie about their true preferences. Unfortunately, the celebrated Gibbard-Satterthwaite theorem~\cite{Gib73,Sat75} shows that if there are at least three possible candidates, then any deterministic strategy-proof voting rule has to be dictatorial---that is, there exists a fixed voter whose top choice is always the winner. Although the Gibbard-Satterthwaite theorem only applies to \emph{deterministic} voting rules, Gibbard later generalized the Gibbard-Satterthwaite theorem to \emph{randomized} voting rules \cite{Gib77}. In particular, Gibbard showed that any randomized strategy-proof voting rule has to be a probability distribution over \emph{unilateral rules} and \emph{duple rules}, where a unilateral rule depends only on a single voter, and a duple rule chooses only between two possible candidates; furthermore, if the voting rule satisfies the natural condition of \emph{Pareto efficiency}---that is, the voting rule \emph{never} chooses a candidate $y$ that is dominated by some other candidate $x$ by every voter---then the voting rule must be a probability distribution over dictatorial voting rules.

The notion of strategy-proofness, however, is quite strong. It requires voters to truthfully report their preferences, \emph{no matter} what preferences the other voters have (and in particular, even if the voter \emph{knows} exactly the preferences of everyone else). One may thus hope that these impossibility results can be circumvented by relaxing this requirement. For instance, for the case of ``large-scale'' voting, it makes sense to assume that each voter has some belief about the preferences of the other voters, and additionally that these preferences are independent and identically distributed (i.i.d.)---we refer to such a notion as \emph{strategy-proofness w.r.t.~(with respect to) i.i.d.~beliefs}. Unfortunately, this weakening does not make things much better: A recent result by McLennan \cite{McL11} shows that if an \emph{anonymous}\footnote{A voting rule is anonymous if it does not depend on the identity of the voters.} voting rule (with at least 3 candidates) is strategy-proof w.r.t.~\emph{all} i.i.d.~beliefs and is also Pareto efficient, then the voting rule must be a random dictatorship---that is, a uniformly random voter's top choice is chosen as the winner. 

Pareto efficiency, however, is a strong condition. When dealing with \emph{randomized} voting rules, a natural relaxation (borrowing from the literature on randomized algorithms or cryptography) would be to allow Pareto efficiency to be violated with some ``tiny'' probability $\eps$. For instance, if this probability $\eps$ is exponentially small in the number of voters, then it seems like the voting rule is still perfectly reasonable. Unfortunately, our first theorem shows that relaxing to $\eps$-Pareto efficiency does not help, even for rather large values of $\eps$, and even for a significantly weaker notion of Pareto efficiency which we call $\eps$-\emph{super-weak unanimity}: $\eps$-super-weak unanimity requires that for every candidate $x$, there exists \emph{some} preference profile with $x$ being the top choice of every voter, such that the voting rule chooses $x$ with probability at least $1-\eps$.

\begin{theorem}[Informal]
Suppose there are at least three candidates. Let $v$ be any anonymous randomized voting rule that is strategy-proof w.r.t.~all i.i.d.~beliefs, and satisfies $\eps$-super-weak unanimity. Then, $v$ is $O(\eps)$-close to the random dictatorship voting rule.
\end{theorem}

Thus, even for an extremely weak notion of what it means to be a ``reasonable'' voting rule, strategy-proofness w.r.t.~all i.i.d.~beliefs cannot be achieved.

\paragraph{Can bounded-rationality help?} In this paper, we consider using notions of ``bounded-rationality'' (see e.g., \cite{Sim55}) to overcome the above impossibility results. An initial approach in this direction was considered by Bartholdi, Tovey, and Trick \cite{BTT89}, who suggested that although voting manipulations exist, they may be ``hard to find''. However, a more recent line of research \cite{Kel93,CS06,PR07,FKN08,FKKN11,XC08,DP08,IKM12,MR12,MR12b} has demonstrated that instances where manipulation is possible are (relatively) common and furthermore, successful manipulation can be efficiently computed (if the manipulator has complete knowledge of everyone else's preferences). Nevertheless, it may still be conceivable that such computational approaches may be applicable if we restrict to strategy-proofness w.r.t.~i.i.d.~beliefs (as in the work of McLennan \cite{McL11}). We do not pursue this path here.

A different approach suggested by Birrell and Pass \cite{BP11} relaxes strategy-proofness to \emph{approximate} strategy-proofness, where a voting rule is $\epsilon$-strategy-proof if no voter can gain more than $\epsilon$ in expected utility by lying. However, although \cite{BP11} presents positive results for the case where $\epsilon = O(1/n)$, they also show that Gibbard's result \cite{Gib77} extends when $\epsilon = o(1/n^2)$. While it may be reasonable to assume that (bounded-rational) voters do not care about ``small'' differences in expected utility, in some settings a gain of $1/n^2$ may be too much. Additionally, Carroll \cite{Car11} demonstrates that a variant of McLennan's result \cite{McL11} holds even if we just consider $o(1/n^{3/2})$-strategy-proofness w.r.t.~i.i.d~beliefs, as long as we restrict to \emph{deterministic} voting rules.

We here consider a new approach to bounded-rationality in voting: we assume that voters have ``coarse'' beliefs.

\paragraph{Strategy-proof voting w.r.t.~coarse i.i.d.~beliefs.} Several celebrated works in the behavioral economics literature (e.g., see \cite{Mul02,MSS08}) indicate that humans ``think through categories'' and that a more appropriate model of human behavior is obtained by restricting players to have ``coarse'' beliefs, where the probabilities are restricted to some coarse set (e.g., a discretization of $[0,1]$) instead of a continuous interval. In this paper, we focus on such ``coarse'' beliefs: we say that a belief is $\alpha$-coarse if the probabilities (the player assigns to states) are restricted to lie on a uniform discretization of $[0,1]$ with ``mesh size'' at least $\alpha$. Coarse beliefs are very natural. For example, any belief with rational probabilities is an $\alpha$-coarse belief for some $\alpha > 0$. Also, many natural methods for forming a belief from observations yield $\alpha$-coarse beliefs where $\alpha$ is inversely proportional to the number of observations; such methods include taking empirical frequencies, as well as using a Dirichlet distribution and updating it when samples or data are observed. We note that even if people form their beliefs using some complicated formula and for instance obtain a belief of the form ``event $A$ happens with probability $1/\sqrt{2}$'', behavioral experiments (see, e.g., \cite{MRM95, MR94}) suggest that people often ``round'' such beliefs and interpret them using some coarse measure (e.g., event $A$ happens with ``very high''/``high''/``medium''/``low''/``very low'' probability).

In this paper, we consider strategy-proofness w.r.t.~coarse i.i.d.~beliefs. We focus on ``large-scale'' voting, where the number of voters $n$ is sufficiently large but is still polynomially-related to $1/\alpha$, where $\alpha$ is the coarseness parameter.

\begin{definition}[Informal] A voting rule is \emph{large-scale strategy-proof w.r.t.~coarse i.i.d.~beliefs} if there exists a polynomial $p(\cdot)$ such that for every coarseness parameter $\alpha > 0$, and every $n \geq p(1/\alpha)$, no voter having an $\alpha$-coarse i.i.d.~belief can improve her expected utility by lying about her preferences.
\end{definition}

We show that there exist anonymous $\eps$-Pareto efficient voting rules that are large-scale strategy-proof w.r.t.~coarse i.i.d.~beliefs, where $\eps$ is exponentially small in the number of voters. 

\begin{theorem}[Informal]
There exist anonymous $\eps$-Pareto efficient voting rules that are large-scale strategy-proof w.r.t.~coarse i.i.d.~beliefs, where $\eps$ is exponentially small in the number of voters.
\end{theorem}

Since we are interested in large-scale voting, where we envision the number of voters to exceed 10000, we do not consider the fact that the voting rule only achieves $\eps$-Pareto efficiency (as opposed to ``exact'' Pareto efficiency) unappealing; the probability of Pareto efficiency being violated is on the order of $2^{-100}$.

\paragraph{Relaxing the coarse i.i.d.~belief assumption.} So far we have assumed that each voter has an $\alpha$-coarse i.i.d.~belief. It is well-known that the i.i.d.~assumption is seemingly strong in the context of voting. To illustrate this, let us recall an example from Chamberlain and Rothschild \cite{CR81}: Consider a simple majority-rule election with two candidates $A$ and $B$. If a voter believes that each of the other voters will vote for candidate $A$ with probability \emph{exactly} $p=0.51$, then in a large-scale election, the voter will be essentially certain that candidate A will win (the probability of him casting the pivotal vote will be on the order of $e^{-n}$, where $n$ is the number of voters). On the other hand, if a voter is \emph{uncertain} about the probability $p$ that the other voters will vote for candidate $A$ (e.g., $p$ is drawn from some distribution over $[0.49, 0.53]$), then this voter may believe that both candidates have a significant chance of winning the election and that the probability of him casting the pivotal vote will be on the order of $1/n$. Note that in the latter case (when the voter is uncertain about $p$), he no longer has an i.i.d.~belief about the preferences of the other voters (conditioned on $p$, the belief is indeed i.i.d., but the combined process of first sampling $p$ and then sampling $n-1$ independent preferences (according to $p$) for the other voters does not result in an i.i.d.~belief; see \cite{CR81} for more discussion on this).

We note, however, that the belief considered above is a \emph{distribution} over i.i.d.~beliefs: we first sample a belief, and then independently sample preferences for the other voters according to this belief. Since our notion of large-scale strategy proofness requires strategy-proofness w.r.t.~\emph{all} coarse i.i.d.~beliefs, it directly follows that our notion implies strategy-proofness w.r.t.~to \emph{all distributions} over coarse i.i.d.~beliefs (e.g., the uniform distribution over a discretization of $[0.49, 0.53]$ in the above example).

Another seemingly strong aspect of i.i.d.~beliefs is that a voter believes that each of the other voters' preferences is drawn from the \emph{same} distribution $\phi$ (e.g., the distribution determined by $p$ in the above example). Again, this assumption can be relaxed by allowing the voter to have a \emph{distribution} over possible $\phi$'s, and a new $\phi$ is sampled for each of the other voters when sampling their preferences. Such a distribution over possible $\phi$'s can be collapsed to a single distribution over preferences. In the case of \emph{coarse} i.i.d.~beliefs, as long as the distribution over possible $\phi$'s has finite support and does not depend on the number of voters, the collapsed distribution will be a \emph{coarse} i.i.d.~belief. Thus, our notion of large-scale strategy proofness w.r.t.~coarse i.i.d.~beliefs also directly implies strategy-proofness in this more complicated model. This more complicated model can be used to model situations where a voter believes that the voting population is separated into a constant number of communities, and each of the communities has a different distribution $\phi$ that is used to generate the community's preferences. For simplicity of presentation, we will state our definitions and results in the more simple model.

\subsection{Our Construction} 

Our construction proceeds in two steps. We first show how to construct voting rules that satisfy exact Pareto efficiency but only a notion of large-scale $\eps$-strategy-proofness w.r.t coarse i.i.d.~beliefs---that is, voters can gain at most $\eps$ in expected utility by lying---where $\eps$ is exponentially small in the number of voters $n$. In a second step, we then show how to transform these voting rules into ones that satisfy actual strategy-proofness w.r.t coarse i.i.d.~beliefs (this, however, comes at the cost of achieving only $\eps$-Pareto efficiency, where $\eps$ is exponentially small).

\paragraph{Step 1: Achieving $\eps$-strategy-proofness.}
We now explain (at a high level) how we obtain voting rules that are large-scale $\eps$-strategy-proof w.r.t.~coarse i.i.d.~beliefs, where $\eps$ is exponentially small. To provide some intuition, let us first consider the plurality rule, which simply chooses the candidate with the most top-choice votes. The plurality rule is not large-scale $\epsilon$-strategy-proof w.r.t.~coarse i.i.d.~beliefs for an exponentially small $\epsilon$. For example, suppose that a voter has the preference ordering $c > a > b$, but she believes that each of the other voters has either the preference ordering $a > b > c$, or the preference ordering $b > a > c$, each with probability $1/2$. Such a belief is $\alpha$-coarse for every $\alpha \leq 1/2$. Now, we observe that according to her belief, her top choice $c$ will certainly not be the winner, so she may want to lie and report her second top choice $a$ as her top choice instead; it can be shown that by doing so, she can increase her expected utility by $\Omega(1/\sqrt{n})$.\footnote{For example, this can be shown by using the analytical tools in \cite{Car11} to establish that with probability $\Omega(1/\sqrt{n})$, the number of top-choice votes for $a$ is equal to that of $b$, in which case the voter can lie to make $a$ the winner.} In this example, the problem is that the voter believes there are two equally popular candidates, namely $a$ and $b$, and by lying, she can make it more likely that the plurality rule chooses candidate $a$ instead of candidate $b$. 

We now show how to modify the plurality rule in (what we consider) a natural way to make it large-scale $\eps$-strategy-proof w.r.t.~coarse i.i.d.~beliefs, where $\eps$ is exponentially small. Recall that each voter submits a preference ordering over the entire set of candidates. Our \emph{Repeated Plurality Elimination} voting rule proceeds as follows: If there is a ``clear winner''---that is, the candidate with the highest number of top-choice votes beats all the other candidates by a \emph{sufficiently large margin} (in terms of the number of top-choice votes), say $\approx \sqrt{n}$, then that candidate is chosen as the winner. Otherwise, we keep all the ``great'' candidates (i.e., the candidates that are within the $\approx \sqrt{n}$ margin of the top candidate) and eliminate all the others. We then restrict the voters' preference orderings to only the ``great'' candidates and recompute the top-choice votes. We repeat this process until either a ``clear winner'' has been found, or no more candidates can be eliminated; in the latter case, we finally apply the traditional plurality rule (without a ``margin'') on the remaining candidates.

Intuitively, these modifications solve the specific issue given above where the voter lies and reports her second top choice $a$ as her top choice, since by Chernoff bounds, with extremely high probability w.r.t.~her belief, the number of top-choice votes for candidates $a$ and $b$ will be within the margin $\approx \sqrt{n}$ of each other; in this case, the voter's lie has no effect, since candidate $c$ will be eliminated while candidates $a$ and $b$ will move onto the next iteration. In our voting rule, the elimination process is repeated because after some candidates are eliminated and the top-choice votes are recounted, the same issue may still be present among the remaining candidates. 

More generally, to prove that our Repeated Plurality Elimination voting rule is large-scale $\eps$-strategy-proof w.r.t.~coarse i.i.d.~beliefs, we intuitively proceed in two steps: In the first step, we show that a voter having a \emph{coarse belief} believes that she only has an exponentially small chance of influencing which candidates get eliminated: very roughly speaking, a candidate's expected count (w.r.t.~to the voter's belief) is either a) \emph{equal to the best candidate's expected count}, in which case by Chernoff bounds, the candidate's actual count will be within the margin (with overwhelming probability), or b) \emph{different from the best candidate's expected count}, in which case, by Chernoff bounds and the fact that the voter's belief is coarse, the candidate's actual count will be outside the margin (with overwhelming probability). In the second step, we show that according to each voter's belief, at the end of the elimination process, we arrive at a situation where all the remaining candidates have \emph{exactly} the same expected count (with overwhelming probability). But if we are in such a situation, then running plurality is actually strategy-proof.

We also consider a variant of the above Repeated Plurality Elimination voting rule, which we refer to as the \emph{approximate instant-runoff} voting rule: at each iteration, instead of eliminating all the candidates that are not ``great'', we instead eliminate all the candidates that are ``close'' to the \emph{worst} candidate, with the following exception: if elimination would cause all the candidates to be eliminated (i.e., all the candidates are close to the worst one), we select the winner using the plurality rule. This voting rule is very similar to the widely used \emph{instant-runoff} voting rule. Instant-runoff voting, as well as variations of it, are used in many elections throughout the world (e.g., see \cite{IRV14}). Instant-runoff voting is identical to our ``approximate instant-runoff'' voting rule with the exception that at each iteration only the candidate with the actual \emph{least} number of top-choice votes is eliminated (as opposed to eliminating all the candidates that are close to it).

More generally, we develop a general framework for constructing voting rules that are large-scale $\eps$-strategy-proof w.r.t.~coarse i.i.d.~beliefs, where $\eps$ is exponentially small, and show how both of the above voting rules (as well as several other voting rules) are natural instances of our framework. All of these voting rules satisfy \emph{exact} Pareto efficiency.

\paragraph{Step 2: Achieving actual strategy-proofness.} In the second step of our construction, we provide a general technique for converting voting rules that are large-scale $\eps$-strategy-proof w.r.t.~coarse i.i.d.~beliefs, where $\eps = o(1/n^2)$, into voting rules that are large-scale (actual) strategy-proof w.r.t.~coarse i.i.d.~beliefs. In fact, such a technique was already provided in \cite{BP11} in the context of strategy-proofness without beliefs (and a variant of it was also explored in \cite{NST12} and \cite{FKKV13} in more general mechanism design contexts), and we here extend the analysis to the context of strategy-proofness with beliefs. The idea (from \cite{BP11}) is to combine in a randomized way an $\eps$-strategy-proof voting rule with a so-called ``punishing'' voting rule that is \emph{strictly} strategy-proof (i.e., voters are strictly better off by truthfully reporting their preferences). The punishing voting rule may not be Pareto efficient, but the combination is done in such a way that the punishing voting rule is only run with tiny probability; this suffices for ensuring that the final voting rule satisfies actual strategy-proofness w.r.t.~coarse i.i.d.~beliefs. Using this technique, we transform our voting rules into ones that satisfy actual strategy-proofness w.r.t coarse i.i.d.~beliefs while satisfying $\eps$-Pareto efficiency, where $\eps$ is exponentially small. This technique actually requires the utility functions of the voters to be coarse, so we will add this assumption to our definition of large-scale strategy-proofness w.r.t. coarse i.i.d.~beliefs; a utility function is $\alpha$-coarse if for every pair of candidates, the utility assigned to the two candidates are either the same or separated by a gap of at least $\alpha$. 

\paragraph{Discussion.} 
If we ignore our use of the punishing voting rule (which is only run with exponentially small probability), our voting rules (e.g., our approximate instant-runoff voting rule) are all quite natural and very similar to what is used in elections throughout the world. Thus, our results provide some intuition for why strategic misreporting of preferences might not be occurring much in these elections.

We note that our voting rules are not monotone---that is, improving the ranking of a candidate in some voter's preference can decrease the chance of that candidate winning. This is because improving the ranking of a candidate in some voter's preference can change which candidates get eliminated, which then changes the number of top-choice votes each candidate has. This side effect can also occur in the classic instant-runoff voting rule, which is also not monotone.

\subsection{Other Related Work}

In this paper, we consider strategy-proofness with respect to a restricted class of beliefs. There have been other papers that also consider strategy-proofness with respect to a restricted class of beliefs. In \cite{MS04}, Majumdar and Sen show that a large class of voting rules are strategy-proof w.r.t.~the uniform belief where the other voters' preferences are uniformly distributed. The authors also show that it is not possible to construct a reasonable deterministic voting rule that is strategy-proof w.r.t.~any of a large set of beliefs where the voters' preferences are independent of each other; this further suggests that the consideration of independent preferences is not sufficient and that it is appropriate to further assume that the preferences are identically distributed. In \cite{She13}, Shen used Beta distributions to model the beliefs of voters (in a way that is different from how we model beliefs) in the context of approval voting, and showed that voters may still have incentives to lie. In contrast to the above two papers---which consider very specific types of beliefs---our focus here is on defining a \emph{general} class of natural beliefs for which strategy-proof voting can be achieved.

\subsection{Outline of Paper}

The rest of the paper is organized as follows. Section \ref{sec:preliminaries} contains the preliminaries. Section \ref{sec:frameworkAndExamples} contains the definition of large-scale strategy-proof w.r.t.~coarse i.i.d.~beliefs, our general framework, and example voting rules. Section \ref{sec:impossibility} contains our impossibility result. All proofs can be found in the appendices.

\section{Preliminaries}
\label{sec:preliminaries}

Given an integer $k \in \N$, let $[k] = \{1, \ldots, k\}$. Let $\C$ be any finite set of \emph{candidates} (or \emph{alternatives}). A \emph{preference ordering} on $\C$ is a strict total order on the set of candidates $\C$; let $\P$ denote the set of all preference orderings on $\C$. Given a subset $A \subseteq \C$ of candidates, let $L(A)$ denote the set of preference orderings (i.e., strict total orders) on $A$. Given a preference ordering $P$ and a pair of candidates $x,y \in \C$, we shall write $x \pref{P} y$ to mean that \emph{$x$ is (strictly) preferred over $y$ in $P$}, i.e., $x$ is ranked higher than $y$ according to $P$. Given a preference ordering $P$, let $top(P)$ denote the highest-ranked candidate according to $P$, i.e., $top(P)$ is the candidate $x$ in $\C$ such that $x \pref{P} y$ for every $y \in \C \setminus \{x\}$. 

Throughout this paper, we will use $n$ to denote the number of voters, and $m$ to denote the number of candidates in $\C$; we will often treat $m$ as a constant. A \emph{preference profile} is a vector of length $n$ whose components are preference orderings in $\P$; that is, a preference profile is simply an element of $\P^n$ which specifies the (submitted) preference orderings of $n$ voters. 

Given a finite set $S$, let $\Delta(S)$ denote the set of all probability distributions over $S$. A (randomized) \emph{voting rule} is a function $v: \P^n \to \Delta(\C)$ (or $v: \P^* \to \Delta(\C)$ if $v$ works for any number of voters) that maps preference profiles to probability distributions over candidates; intuitively, $v(\vec{P})$ is a distribution over $\C$ that specifies the probability that each candidate is selected when the submitted votes form the preference profile $\vec{P}$. A voting rule $v$ is said to be \emph{deterministic} if for every preference profile $\vec{P}$, the distribution $v(\vec{P})$ assigns probability 1 to some candidate. A voting rule $v$ is said to be \emph{anonymous} if $v$ does not depend on the order in which the preference orderings appear in the input, i.e., $v(P_1, \ldots, P_n) = v(P_{\sigma(1)}, \ldots, P_{\sigma(n)})$ for every preference profile $(P_1, \ldots, P_n) \in \P^n$ and every permutation $\sigma: [n] \to [n]$. In this paper, we will only consider anonymous voting rules; most common voting rules are indeed anonymous, and one can argue that anonymous voting rules are more fair and democratic than non-anonymous ones. 

Given a (randomized) voting rule $v: \P^n \to \Delta(\C)$, a candidate $x \in \C$, and a preference profile $\vec{P}$, let $v(x,\vec{P})$ be the probability mass assigned to $x$ by the distribution $v(\vec{P})$; we also refer to $v(x,\vec{P})$ as the \emph{selection probability of $x$ with respect to $v$ and $\vec{P}$}, since $v(x,\vec{P})$ is the probability that candidate $x$ is selected by the voting rule $v$ when the input preference profile is $\vec{P}$. A \emph{utility function} is a function $u: \C \to [0,1]$ that assigns a real number in $[0,1]$ to each candidate in $\C$ \footnote{It is not important that the codomain of the utility function $u$ is $[0,1]$; as long as the codomain is bounded, the results of this paper still hold with minor modifications.}. Given a preference ordering $P$ and a utility function $u$, we say that $u$ is \emph{consistent with $P$} if for every pair of candidates $x,y \in \C$, we have $u(x) > u(y)$ if and only if $x \pref{P} y$.

A voting rule is \emph{Pareto efficient} if it never chooses a Pareto dominated candidate, i.e., a candidate $y$ such that all the voters prefer $x$ over $y$ for some candidate $x$. A slight relaxation of Pareto efficiency is \emph{$\eps$-Pareto efficiency}, where we allow the voting rule to choose a Pareto dominated candidate with probability at most $\eps$.

\begin{definition}[$\eps$-Pareto efficiency]
A voting rule $v: \P^n \to \Delta(\C)$ is \emph{$\eps$-Pareto efficient} if for every pair of candidates $x,y \in \C$ and every preference profile $\vec{P} = (P_1, \ldots, P_n) \in \P^n$ such that $x \pref{P_i} y$ for every $i \in [n]$, we have $v(y,\vec{P}) \leq \eps$. 
\end{definition}

The \emph{random dictatorship} voting rule is the voting rule $v_{dict}$ that chooses a voter uniformly at random and then chooses her top choice, i.e., for every preference profile $\vec{P} = (P_1, \ldots, P_n) \in \P^n$ and every candidate $x \in \C$, we have $v_{dict}(x,\vec{P}) = \frac{|\{i \in [n] : top(P_i) = x\}|}{n}$. 

See \ref{app:Gibbard-Satterthwaite} for background information on the Gibbard-Satterthwaite theorem \cite{Gib73,Sat75} and Gibbard's generalization of the Gibbard-Satterthwaite theorem to randomized voting rules \cite{Gib77}.

\subsection{Strategy-Proofness with respect to a Set of Beliefs} 

Gibbard's generalization \cite{Gib77} of the Gibbard-Satterthwaite theorem shows that when there are at least three candidates, we cannot even construct good \emph{randomized} voting rules that are strategy-proof. Given this impossibility result, let us consider relaxed notions of strategy-proofness. We observe that strategy-proofness requires that no voter would want to lie about her true preference even if the voter \emph{knows} the submitted preferences of \emph{all} the other voters. However, in many realistic scenarios, a voter is \emph{uncertain} about how other voters will vote, and she would only lie if she \emph{believes} that she can gain utility in expectation by lying. As a result, we consider a relaxed notion of strategy-proofness where we consider the voter's \emph{belief} of how the other voters will vote. The standard notion of strategy-proofness requires that no voter would want to lie regardless of what her belief is. To weaken the notion of strategy-proof, one can require that no voter would want to lie as long as her belief belongs in a certain set of beliefs. Let us now move to formalizing these notions.

In this paper, we will only consider beliefs that are \emph{i.i.d.}~(independent and identically distributed), meaning that for each belief, the other voters' preference orderings are sampled independently from some distribution $\phi$ over preference orderings. Thus, for simplicity, we define a \emph{belief} to be a probability distribution over the set $\P$ of preference orderings, representing a voter's belief that each of the other voters will have a preference ordering drawn independently from this distribution. We now state the definition of \emph{strategy-proof with respect to a set of beliefs}.

\begin{definition}[Strategy-proof w.r.t.~a set $\Phi$ of beliefs]
A voting rule $v: \P^n \to \Delta(\C)$ is \emph{strategy-proof w.r.t.~a set $\Phi$ of beliefs} if for every $i \in [n]$, every pair of preference orderings $P_i, P_i' \in \P$, every belief $\phi \in \Phi$, and every utility function $u_i$ that is consistent with $P_i$, we have
\begin{align*}
  \E[u_i(v(\vec{P}_{-i},P_i))] \geq \E[u_i(v(\vec{P}_{-i},P_i'))],
\end{align*}
where $\vec{P}_{-i} \sim \phi^{n-1}$.
\end{definition}

\section{Large-Scale Strategy-Proof Voting w.r.t.~Coarse i.i.d.~Beliefs}
\label{sec:frameworkAndExamples}

In this section, we first define the notion of ``coarse'' i.i.d.~beliefs; then, we introduce the notion of \emph{large-scale strategy-proof w.r.t.~coarse i.i.d.~beliefs}. We then develop a general framework for constructing voting rules that are large-scale $\eps$-strategy-proof w.r.t.~coarse i.i.d.~beliefs, and we then use the general framework to obtain many examples of good voting rules. We then show how to transform these voting rules into ones that are \emph{actually} large-scale strategy-proof w.r.t.~coarse i.i.d.~beliefs.

Let us begin by introducing the notion of a coarse i.i.d.~belief. Roughly speaking, an i.i.d.~belief $\phi$ is $\alpha$-coarse if the probability masses assigned by $\phi$ are restricted to lie on a uniform discretization of $[0,1]$ with ``mesh size'' at least $\alpha$. More precisely, an i.i.d.~belief $\phi$ is said to be \emph{$\alpha$-coarse} if the probability masses assigned by $\phi$ are multiples of some number $\beta \geq \alpha$, i.e., there exists a number $\beta \geq \alpha$ such that for every preference ordering $P \in \P$, we have $\phi(P) = i\beta$ for some integer $i$. Coarse i.i.d.~beliefs are quite natural due to many reasons. For example, if a human were to describe or represent her belief (as a distribution over preference orderings), the probabilities would almost certainly be rational numbers (e.g., it is very strange to believe that a certain preference ordering has probability $1/\pi$ of occurring), and an i.i.d.~belief with rational probabilities is an $\alpha$-coarse i.i.d.~belief for some $\alpha > 0$. Also, many common and natural ways of forming a belief also result in a coarse i.i.d.~belief. For example, one can use empirical frequencies or a Dirichlet distribution to form a belief from observed samples of preferences. Both of these methods yield $\alpha$-coarse i.i.d.~beliefs, where $\alpha$ is inversely proportional to the number of observations. See \ref{app:formingBeliefs} for more information. We can also consider $\alpha$-coarse utility functions. A utility function $u: \C \to [0,1]$ is said to be \emph{$\alpha$-coarse} if for every pair of candidates $x,y \in \C$, we have $u(x) = u(y)$ or $|u(x) - u(y)| \geq \alpha$. We only need the utility functions to be coarse for the ``punishing'' voting rule that we will use later.

\paragraph{Large-scale strategy-proof w.r.t.~coarse i.i.d.~beliefs.} Let us now introduce the notion of \emph{large-scale strategy-proof w.r.t.~coarse i.i.d.~beliefs}, which is a notion of strategy-proof where the voters have coarse i.i.d.~beliefs and there are sufficiently (but still polynomially) many voters.

\begin{definition}[Large-scale strategy-proof w.r.t.~coarse i.i.d.~beliefs]
A voting rule $v: \P^* \to \Delta(\C)$ is \emph{large-scale strategy-proof w.r.t.~coarse i.i.d.~beliefs} if there exists a polynomial $p(\cdot)$ such that for every $\alpha > 0$, every $n \geq p(\frac{1}{\alpha})$, every $i \in [n]$, every pair of preference orderings $P_i, P_i' \in \P$, every $\alpha$-coarse i.i.d.~belief $\phi_i$, and every $\alpha$-coarse utility function $u_i$ that is consistent with $P_i$, we have
\begin{align*}
  \E[u_i(v(\vec{P}_{-i},P_i))] \geq \E[u_i(v(\vec{P}_{-i},P_i'))],
\end{align*}
where $\vec{P}_{-i} \sim {\phi_i}^{n-1}$; when this holds, we may refer to the above polynomial $p(\cdot)$ as the \emph{rate} of the voting rule $v$.
\end{definition}

In the above definition, $\alpha$ controls the coarseness of the belief, and $p(1/\alpha)$ controls how many voters are required in order to achieve truthfulness; we need $n$ to be sufficiently large because as the i.i.d.~beliefs become less and less coarse, the set of beliefs considered becomes closer and closer to the set of all i.i.d.~beliefs, which we later show is impossible to construct good voting rules for. The rate $p(\cdot)$ captures how many voters are needed relative to the coarseness of the beliefs. 

As mentioned in the introduction, we can consider a slightly more realistic model where each voter has a distribution over $\alpha$-coarse i.i.d.~beliefs, and when computing expected utility for a voter, a \emph{single} $\alpha$-coarse i.i.d.~belief is sampled from this distribution, and then this sampled belief is used to generate \emph{all} the other voters' preferences in an i.i.d.~manner. Our results still hold in this more realistic model; this easily follows from the definition of large-scale strategy-proof w.r.t.~coarse i.i.d.~beliefs, where it is required that strategy-proofness holds for \emph{every} $\alpha$-coarse belief, so strategy-proofness also holds if we sample a random $\alpha$-coarse i.i.d.~belief from a distribution.

We now define a relaxed version of large-scale strategy-proof w.r.t.~coarse i.i.d.~beliefs, where we allow a voter to gain at most $\eps(n)$ in expected utility.

\begin{definition}[Large-scale $\eps$-strategy-proof w.r.t.~coarse i.i.d.~beliefs]
A voting rule $v: \P^* \to \Delta(\C)$ is \emph{large-scale $\eps$-strategy-proof w.r.t.~coarse i.i.d.~beliefs} if there exists a polynomial $p(\cdot)$ such that for every $\alpha > 0$, every $n \geq p(\frac{1}{\alpha})$, every $i \in [n]$, every pair of preference orderings $P_i, P_i' \in \P$, every $\alpha$-coarse i.i.d.~belief $\phi_i$, and every $\alpha$-coarse utility function $u_i$ that is consistent with $P_i$, we have
\begin{align*}
  \E[u_i(v(\vec{P}_{-i},P_i))] \geq \E[u_i(v(\vec{P}_{-i},P_i'))] - \eps(n),
\end{align*}
where $\vec{P}_{-i} \sim {\phi_i}^{n-1}$; when this holds, we may refer to the above polynomial $p(\cdot)$ as the \emph{rate} of the voting rule $v$.
\end{definition}

\subsection{Our General Framework}
\label{sec:framework}

In this section, we develop a general framework for constructing large-scale $\eps$-strategy-proof voting rules w.r.t.~coarse i.i.d.~beliefs. Later, we will show how to transform such voting rules into ones that satisfy \emph{actual} large-scale strategy-proofness w.r.t coarse i.i.d.~beliefs. Before we describe the general framework in detail, let us describe an example for motivation. 

Recall that the plurality rule simply chooses the candidate with the most top-choice votes. The plurality rule is simple, very commonly used, and intuitively has good efficiency (e.g., it is Pareto efficient). Unfortunately, the plurality rule is not large-scale $\eps$-strategy-proof w.r.t.~coarse i.i.d.~beliefs for any exponentially small $\eps$. However, it is not hard to see that the plurality rule is strategy-proof w.r.t.~beliefs where all the candidates have the same probability of being the top choice of a voter's preference ordering. Can one design an ``elimination rule'' that eliminates candidates in a way so that (1) a voter with a coarse i.i.d.~belief will believe that she only has an exponentially small chance of affecting which candidates get eliminated, and (2) once these candidates are eliminated from her belief, all the remaining candidates will have the same probability of being the top choice? Intuitively, by running such an elimination rule and then running the plurality rule on the remaining candidates, the combined voting rule is large-scale $\eps$-strategy-proof w.r.t.~coarse i.i.d.~beliefs, where $\eps$ is exponentially small. Such an elimination rule exists; it repeatedly eliminates the candidates whose number of top-choice votes is not ``close'' to the highest number of top-choice votes. We will later show that this elimination rule satisfies the two required properties.

The above example can be viewed as an instantiation of a more general framework for constructing voting rules that are large-scale $\eps$-strategy-proof w.r.t.~coarse i.i.d.~beliefs, which we now describe. The general framework consists of an ``elimination rule'' and a ``selection rule'' satisfying certain properties. The elimination rule will choose a subset of the candidates, and then the selection rule will select a winner from this subset. As long as certain properties are satisfied, the elimination rule combined with the selection rule will be large-scale $\eps$-strategy-proof w.r.t.~coarse i.i.d.~beliefs. Let us now informally describe the general procedure and the requirements.

On input a preference profile, we do the following:

\begin{description}
	\item[Stage 1:] Run an ``elimination rule'' that, on input a preference profile, eliminates a subset of the candidates, leaving a subset $A \subseteq \C$ remaining. We require the following: a single voter with a coarse i.i.d.~belief $\phi$ has little influence on the choice $A$ of the elimination rule when the other voters' preferences are distributed according to $\phi$; furthermore, with high probability, the restriction of the belief $\phi$ to the remaining candidates $A$ results in a belief in some set $\Phi'_A$. 
	
	\item[Stage 2:] Run a ``selection rule'' on the preference profile restricted to the remaining candidates $A$. We require that the selection rule is strategy-proof w.r.t.~the set $\Phi'_A$ of beliefs from Stage 1. 
\end{description}

Intuitively, the above procedure is large-scale $\eps$-strategy-proof w.r.t.~coarse i.i.d.~beliefs because a voter $i$ with a coarse i.i.d.~belief $\phi$ will believe that she has little influence on the choice $A$ of the elimination rule, and since the restriction of $\phi$ to $A$ is a belief for which the selection rule is strategy-proof, voter $i$ cannot gain much by lying. 

We now describe the framework more formally. An \emph{elimination rule} is a function $f: \P^* \to \Delta(2^{\C})$ that, on input a preference profile $\vec{P}$, outputs a non-empty subset $A \subseteq \C$ representing the \emph{remaining} candidates after elimination. Recall that given a subset $A \subseteq \C$ of candidates, we use $L(A)$ to denote the set of preference orderings on $A$. A \emph{selection rule} is a collection of functions $\{s_A: (L(A))^* \to \Delta(A)\}_{A \subseteq \C, A \neq \emptyset}$, one for each non-empty subset $A \subseteq \C$, such that for every $A \subseteq \C$, $s_A$ is a voting rule for the set of candidates $A$. Given a selection rule $s = \{s_A\}_{A \subseteq \C, A \neq \emptyset}$ and a preference profile $\vec{P}$ whose components are preference orderings over $A$, let $s(\vec{P}) = s_A(\vec{P})$. 

Given a preference profile $\vec{P}$ and a non-empty subset $A \subseteq \C$ of candidates, let the \emph{restriction of $\vec{P}$ to $A$}, denoted $\vec{P}|_A$, be the preference profile obtained by removing all the candidates in $\vec{P}$ that are not in $A$, while preserving the ordering of the remaining candidates. Given an i.i.d.~belief $\phi$ and a non-empty subset $A \subseteq \C$ of candidates, let the \emph{restriction of $\phi$ to $A$}, denoted $\phi|_A$, be the belief (i.e., distribution over preference orderings on $A$) $P|_A$, where $P \sim \phi$. We now state our theorem that precisely describes our general framework.

\begin{theorem}[Our general framework]
\label{thm:framework}
Let $f: \P^* \to \Delta(2^{\C})$ be any elimination rule, and let $s = \{s_A\}_{A \subseteq \C, A \neq \emptyset}$ be any selection rule. Let $\delta : \mathbb{N} \to \R$ be any function. Suppose there exists a polynomial $p(\cdot)$ such that for every $\alpha > 0$ and every $n \geq p(\frac{1}{\alpha})$, the following holds: 
\begin{itemize} 
	\item For every $i \in [n]$ and every $\alpha$-coarse i.i.d.~belief $\phi_i$, there exists a non-empty subset $A \subseteq \C$ of candidates such that the following conditions hold:
	\begin{itemize}
		\item For every $P_i \in \P$, the elimination rule $f(\vec{P}_{-i}, P_i)$ chooses (to keep) $A$ with probability at least $1 - \delta(n)$ over the randomness of $\vec{P}_{-i} \sim {\phi_i}^{n-1}$ and $f$.
		\item $s_A$ is strategy-proof w.r.t.~the restricted belief $\phi_i|_A$. 
	\end{itemize}
\end{itemize}
Then, the voting rule $v: \P^* \to \Delta(\C)$ defined by $v(\vec{P}) = s(\vec{P}|_{f(\vec{P})})$ is large-scale $2\delta$-strategy-proof w.r.t.~coarse i.i.d.~beliefs, and the rate of $v$ is the polynomial $p(\cdot)$.
\end{theorem}

See \ref{app:ProofsFromFrameworkAndExamples} for the proof of Theorem \ref{thm:framework}.

\subsection{Examples of our General Framework}
\label{sec:examples}

In this section, we provide some examples of our general framework. Recall that the plurality rule simply chooses the candidate with the most top-choice votes (breaking ties in some way). We now describe a modified plurality rule in the format of our general framework; this voting rule is equivalent to the repeated plurality elimination voting rule described earlier and in the introduction of this paper.

\begin{example}[{\bf Repeated Plurality Elimination + Plurality Selection}]
Let $0 < \delta < 1/2$, and let $v_{pl}: \P^* \to \Delta(\C)$ be a voting rule defined as follows; on input a preference profile $\vec{P} \in \P^n$, $v_{pl}$ does the following:
\begin{description}
	\item[Stage 1:] Repeatedly do the following until no more candidates are eliminated: count the number of top-choice votes for each candidate, and eliminate all the candidates that have a count that is not within $n^{1/2 + \delta}$ of the highest count among the remaining candidates; restrict the preference profile to the set of remaining candidates.
  \item[Stage 2:] Run the plurality rule for the remaining candidates, i.e., on the preference profile restricted to the set of remaining candidates. 
\end{description}

Using our general framework, we now show that $v_{pl}$ is large-scale $\eps$-strategy-proof w.r.t.~coarse i.i.d.~beliefs (where $\eps$ is exponentially small), and also satisfies certain efficiency properties.

\begin{theorem}
\label{thm:v_pl}
Let $0 < \delta < 1/2$, and let $v_{pl}$ be the voting rule defined above. Then, $v_{pl}$ satisfies the following properties:
\begin{enumerate}
	\item $v_{pl}$ is large-scale $(e^{-\Omega(n^{2\delta})})$-strategy-proof w.r.t.~coarse i.i.d.~beliefs, with rate $p(x) = O(x^{\lceil 1/(1/2-\delta) \rceil})$.
	\item $v_{pl}$ is Pareto efficient.
	\item $v_{pl}$ is $n^{1/2+\delta}$-close to optimal in the sense that $v_{pl}$ always chooses a candidate $c \in \C$ such that the number of top-choice votes for $c$ is within $n^{1/2+\delta}$ of the highest number of top-choice votes among the candidates.
\end{enumerate}
\end{theorem}

See \ref{app:ProofsFromFrameworkAndExamples} for the proof of Theorem \ref{thm:v_pl}. Recall that we will later combine this voting rule with a ``punishing'' voting rule to obtain a voting rule that is large-scale (actual) strategy-proof w.r.t.~coarse i.i.d.~beliefs.
\end{example}

\begin{example}[{\bf Approximate Instant-Runoff Voting}]
The standard instant-runoff voting rule repeats the following until a candidate has been chosen as the winner: count the number of top-choice votes for each candidate, and eliminate the candidate with the least number of top-choice votes (breaking ties in some way); restrict the preference profile to the set of remaining candidates, and if there is only one candidate remaining, choose the candidate to be the winner.

It is not hard to see that the standard instant-runoff voting rule is not large-scale $\eps$-strategy-proof w.r.t.~coarse i.i.d.~beliefs, where $\eps$ is reasonably small. However, we can slightly modify the standard instant-runoff voting rule to obtain an approximate version that is large-scale $\eps$-strategy-proof w.r.t.~coarse i.i.d.~beliefs, where $\eps$ is exponentially small. In each iteration, instead of eliminating only the candidate with the least number of top-choice votes, we eliminate all the candidates that have a count that is close to the least number of top-choice votes; however, we stop right before all the remaining candidates are about to be eliminated, and then we choose the candidate with the most top-choice votes. Let us now put our approximate instant-runoff voting rule in the format of our general framework. 

Let $0 < \delta < 1/2$, and let $v_{irv}: \P^* \to \Delta(\C)$ be a voting rule defined as follows; on input a preference profile $\vec{P} \in \P^n$, $v_{irv}$ does the following:
\begin{description}
	\item[Stage 1:] Repeat the following: Count the number of top-choice votes for each candidate, and eliminate all the candidates that have a count that is within $n^{1/2 + \delta}$ of the least number of top-choice votes, unless doing so would eliminate all the remaining candidates, in which case we simply stop and proceed to Stage 2; restrict the preference profile to the set of remaining candidates.
  \item[Stage 2:] Run the plurality rule for the remaining candidates, i.e., on the preference profile restricted to the set of remaining candidates. 
\end{description}
Using our general framework, we now show that $v_{irv}$ is large-scale $\eps$-strategy-proof w.r.t.~coarse i.i.d.~beliefs (where $\eps$ is exponentially small), and also satisfies certain efficiency properties.

\begin{theorem}
\label{thm:v_irv}
Let $0 < \delta < 1/2$, and let $v_{irv}$ be the voting rule defined above. Then, $v_{irv}$ satisfies the following properties:
\begin{enumerate}
	\item $v_{irv}$ is large-scale $(e^{-\Omega(n^{2\delta})})$-strategy-proof w.r.t.~coarse i.i.d.~beliefs, with rate $p(x) = O(x^{\lceil 1/(1/2-\delta) \rceil})$.
	\item $v_{irv}$ is Pareto efficient.
\end{enumerate}
\end{theorem}

See \ref{app:ProofsFromFrameworkAndExamples} for the proof of Theorem \ref{thm:v_irv}. Recall that we will later combine this voting rule with a ``punishing'' voting rule to obtain a voting rule that is large-scale (actual) strategy-proof w.r.t.~coarse i.i.d.~beliefs.
\end{example}

In \ref{app:moreExamples}, we provide some more examples of our general framework.

\subsection{Achieving Actual Strategy-Proofness via the Punishing Voting Rule}
\label{sec:punishingVotingRule}

In this section, we show how to transform our voting rules into ones that are actually strategy-proof w.r.t.~coarse i.i.d.~beliefs. We do this by providing a general technique for converting voting rules that are large-scale $\eps$-strategy-proof w.r.t.~coarse i.i.d.~beliefs, where $\eps = o(1/n^2)$, into voting rules that are large-scale (actual) strategy-proof w.r.t.~coarse i.i.d.~beliefs. The idea is to combine in a randomized way an $\eps$-strategy-proof voting rule with a ``punishing'' voting rule that is ``strictly strategy-proof''. The punishing voting rule is defined as follows:
\begin{itemize}
  \item Let $v_{punish}: \P^n \to \Delta(\C)$ be the voting rule that chooses a voter $i \in [n]$ uniformly at random and then chooses the $j^{th}$ top choice of voter $i$ with probability proportional to $m-j$, i.e., with probability $(m-j) / \sum_{\ell=1}^m (m-\ell)$. 
\end{itemize}
We now show that $v_{punish}$ is \emph{strictly} strategy-proof in the sense that if a voter lies about her preference ordering, her expected utility will be strictly less than what it would be if she submitted her true preference ordering, and the difference in the two expected utilities is at least $\Omega(\alpha/n)$, where $\alpha$ is the coarseness of the utility function.

\begin{lemma}
\label{lem:strictlySP}
The voting rule $v_{punish}$ is ``strictly strategy-proof'' in the following sense: For every $\alpha > 0$, every $i \in [n]$, every pair of preference orderings $P_i, P_i' \in \P$ with $P_i \neq P_i'$, every $\vec{P}_{-i} \in \P^{n-1}$, and every $\alpha$-coarse utility function $u_i$ that is consistent with $P_i$, we have
\begin{align*}
  \E[u_i(v_{punish}(\vec{P}_{-i},P_i))] \geq \E[u_i(v_{punish}(\vec{P}_{-i},P_i'))] + \Omega(\alpha / n).
\end{align*}
\end{lemma}

See \ref{app:punishingVotingRule} for the proof of Lemma \ref{lem:strictlySP}. We now show that if we take a voting rule $v$ that is large-scale $\eps$-strategy proof w.r.t.~coarse i.i.d.~beliefs, where $\eps = o(1/n^2)$, and ``mix'' it with the punishing voting rule $v_{punish}$ by running $v$ with probability $1-q$ and $v_{punish}$ with probability $q$ for some appropriately chosen $q = \Omega(n^2 \cdot \eps(n))$, then the ``mixed'' voting rule is large-scale (actual) strategy-proof w.r.t.~coarse i.i.d.~beliefs.

\begin{lemma}
\label{lem:mixIsSP}
Let $v: \P^* \to \Delta(\C)$ be any voting rule that is large-scale $\eps$-strategy-proof w.r.t.~coarse i.i.d.~beliefs, where $\eps(n) = o(1/n^2)$. Let $v_{mix}$ be the voting rule that runs $v$ with probability $1-q(n)$ and runs $v_{punish}$ with probability $q(n)$, where $q(n) = \Omega(n^2 \cdot \eps(n))$. Then, $v_{mix}$ is large-scale strategy-proof w.r.t.~coarse i.i.d.~beliefs.
\end{lemma}

See \ref{app:punishingVotingRule} for the proof of Lemma \ref{lem:mixIsSP}. We now combine the punishing voting rule with our general framework (Theorem \ref{thm:framework}) to obtain a new general framework for obtaining voting rules that are large-scale (actual) strategy-proof w.r.t.~coarse i.i.d.~beliefs.

\begin{theorem}
\label{thm:fullFramework}
Let $v: \P^* \to \Delta(\C)$ be a voting rule as defined in Theorem \ref{thm:framework} with corresponding function $\delta(n) = o(1/n^2)$. Let $v_{mix}$ be the voting rule that runs $v$ with probability $1-q(n)$ and runs $v_{punish}$ with probability $q(n)$, where $q(n) = \Omega(n^2 \cdot \delta(n))$. Then, $v_{mix}$ is large-scale strategy-proof w.r.t.~coarse i.i.d.~beliefs.
\end{theorem}

\begin{proof}
The theorem follows by combining Theorem \ref{thm:framework} with Lemma \ref{lem:mixIsSP}.
\end{proof}

Using the punishing voting rule, we can also transform our previous voting rules into ones that are large-scale (actual) strategy-proof w.r.t.~coarse i.i.d.~beliefs, and $\eps$-Pareto efficient, where $\eps$ is exponentially small.

\begin{theorem}[{\bf Repeated Plurality Elimination + Plurality Selection}]
\label{thm:pluralityRuleWithPunishing}
There exists a constant $C > 0$ such that the following holds. Let $0 < \delta < 1/2$, and let $v_{pl}$ be the voting rule in Theorem \ref{thm:v_pl}. Let $v_{pl}'$ be the voting rule that runs $v_{pl}$ with probability $1-e^{-Cn^{2\delta}}$ and runs $v_{punish}$ with probability $e^{-Cn^{2\delta}}$. Then, $v_{pl}'$ satisfies the following properties:
\begin{enumerate}
	\item $v_{pl}'$ is large-scale strategy-proof w.r.t.~coarse i.i.d.~beliefs, with rate $p(x) = O(x^{\lceil 1/(1/2-\delta) \rceil})$.
	\item $v_{pl}'$ is $e^{-\Omega(n^{2\delta})}$-Pareto efficient.
	\item With probability at least $1 - e^{-\Omega(n^{2\delta})}$, $v_{pl}'$ is $n^{1/2+\delta}$-close to optimal in the sense that $v_{pl}'$ chooses a candidate $c \in \C$ such that the number of top-choice votes for $c$ is within $n^{1/2+\delta}$ of the highest number of top-choice votes among the candidates.
\end{enumerate}
\end{theorem}

\begin{proof}
The theorem immediately follows by combining Theorem \ref{thm:v_pl} with Lemma \ref{lem:mixIsSP}.
\end{proof}

\begin{theorem}[{\bf Approximate Instant-Runoff Voting}]
\label{thm:v_irvWithPunishing}
There exists a constant $C > 0$ such that the following holds. Let $0 < \delta < 1/2$, and let $v_{irv}$ be the voting rule in Theorem \ref{thm:v_irv}. Let $v_{irv}'$ be the voting rule that runs $v_{irv}$ with probability $1-e^{-Cn^{2\delta}}$ and runs $v_{punish}$ with probability $e^{-Cn^{2\delta}}$. Then, $v_{irv}'$ satisfies the following properties:
\begin{enumerate}
	\item $v_{irv}'$ is large-scale strategy-proof w.r.t.~coarse i.i.d.~beliefs, with rate $p(x) = O(x^{\lceil 1/(1/2-\delta) \rceil})$.
	\item $v_{irv}'$ is $e^{-\Omega(n^{2\delta})}$-Pareto efficient.
\end{enumerate}
\end{theorem}

\begin{proof}
The theorem immediately follows by combining Theorem \ref{thm:v_irv} with Lemma \ref{lem:mixIsSP}.
\end{proof}

\section{Impossibility of Strategy-Proofness w.r.t.~all i.i.d.~Beliefs}
\label{sec:impossibility}

In this section, we prove our impossibility result that says that if there are at least three candidates, then it is not possible to construct a voting rule that is strategy-proof for all i.i.d.~beliefs and satisfies $\eps$-super-weak unanimity, unless the voting rule is $O(\eps)$-close to being the random dictatorship voting rule. This section is self-contained, and the other sections do not rely on the material in this section; as a result, the reader can safely skip this section if he or she wishes to do so.

We begin with some definitions. We call a voting rule \emph{weakly strategy-proof} if it is strategy-proof with respect to the set of all i.i.d.~beliefs.

\begin{definition}[Weakly strategy-proof]
A voting rule $v: \P^n \to \Delta(\C)$ is \emph{weakly strategy-proof} if $v$ is strategy proof with respect to the set of all i.i.d.~beliefs.
\end{definition}

Suppose all the voters have the same top choice, say candidate $x$; then, we would expect the voting rule to choose $x$ as the winner. We call this property \emph{strong unanimity}; again, we slightly relax this property to \emph{$\eps$-strong unanimity}, where we allow the voting rule to choose the common top candidate with probability at least $1-\eps$ instead of probability $1$. 

\begin{definition}[$\eps$-strong unanimity]
A voting rule $v: \P^n \to \Delta(\C)$ satisfies \emph{$\eps$-strong unanimity} if for every candidate $x \in \C$ and every preference profile $\vec{P} = (P_1, \ldots, P_n) \in \P^n$ such that $top(P_i) = x$ for every $i \in [n]$, we have $v(x,\vec{P}) \geq 1 - \eps$. 
\end{definition}

It is easy to see that strong unanimity is weaker than Pareto efficiency (modulo a factor of $m$ for the $\eps$ version of the properties). Now, suppose all the voters have the exact same preference ordering $P$; then, we would expect the voting rule to choose the top candidate of $P$. We call this property \emph{weak unanimity}; again, we state an $\eps$ version of the definition.

\begin{definition}[$\eps$-weak unanimity]
A voting rule $v: \P^n \to \Delta(\C)$ satisfies \emph{$\eps$-weak unanimity} if for every preference ordering $P \in \P$ and every preference profile $\vec{P} = (P_1, \ldots, P_n) \in \P^n$ such that $P_i = P$ for every $i \in [n]$, we have $v(top(P),\vec{P}) \geq 1 - \eps$. 
\end{definition}

It is clear that $\eps$-weak unanimity is weaker than $\eps$-strong unanimity. We finally define \emph{$\eps$-super-weak unanimity}, which is \emph{even weaker} than $\eps$-weak unanimity, and requires that for every candidate $x \in C$, there exists some preference profile $\vec{P}$ with $x$ at the top of every preference ordering in $\vec{P}$, such that the voting rule on $\vec{P}$ will choose $x$ as the winner with probability at least $1-\eps$. 

\begin{definition}[$\eps$-super-weak unanimity]
A voting rule $v: \P^n \to \Delta(\C)$ satisfies \emph{$\eps$-super-weak unanimity} if for every candidate $x \in C$, there exists a preference profile $\vec{P} = (P_1, \ldots, P_n) \in \P^n$ with $top(P_i) = x$ for every $i \in [n]$, such that $v(x, \vec{P}) \geq 1 - \eps$. 
\end{definition}

$\eps$-super-weak unanimity is a very weak property that all reasonable voting rules should have. We now define what it means for two voting rules to be \emph{$\eps$-close} to each other. 

\begin{definition}
Let $v,v': \P^n \to \Delta(\C)$ be two randomized voting rules. We say that $v$ is \emph{$\eps$-close} to $v'$ if for every preference profile $\vec{P}$ and every candidate $x \in \C$, we have $|v(x,\vec{P}) - v'(x,\vec{P})| \leq \eps$. 
\end{definition}

We now formally state our theorem.

\begin{theorem}
\label{thm:impossibility}
Suppose there are at least three candidates in $\C$, i.e., $|\C| \geq 3$. Let $v: \P^n \to \Delta(\C)$ be any anonymous randomized voting rule that is weakly strategy-proof and satisfies $\eps$-super-weak unanimity. Then, $v$ is $O(\eps)$-close to the random dictatorship voting rule.
\end{theorem}

Let us briefly mention some aspects of our proof. Our proof uses some tools from McLennan's impossibility result \cite{McL11} and Gibbard's generalization of the Gibbard-Satterthwaite theorem \cite{Gib77}. However, our result does not follow from generalizing McLennan's proof. For example, using McLennan's proof, it is not clear how one can weaken the assumption of Pareto efficiency to $\eps$-Pareto efficiency and still show that the voting rule is close to random dictatorship. By adding ``error terms'' at various places of McLennan's proof, it is not too difficult to show that the voting rule $v$ is $O(\eps n)$-close to random dictatorship. However, if $\eps n$ is large (e.g., $\eps n = \Omega(1)$), then the result is not meaningful. In particular, such a result does not prevent the possibility of constructing a voting rule that is $1/n$-Pareto efficient and weakly strategy-proof. Our impossibility result shows that the voting rule $v$ is $O(\eps)$-close to random dictatorship, as opposed to $O(\eps n)$-close; in order to prove this, we needed to use different analyses and go through substantially more work.

We will now prove Theorem \ref{thm:impossibility}. We will prove a sequence of lemmas and claims that describe properties that the voting rule $v$ must have, which will be used to show that $v$ is $O(\eps)$-close to the random dictatorship voting rule.

We first establish some notation and terminology that will be used later; most of the notation and terminology comes from \cite{Gib77}, which is Gibbard's generalization of the Gibbard-Satterthwaite theorem to randomized voting rules. Given a preference ordering $P$ and a pair of candidates $x, y \in \C$, we shall write $x \pref{P!} y$ to mean that $x$ is directly on top of $y$ in the preference ordering $P$, i.e., $x \pref{P} y$ and for every candidate $z \in \C \setminus \{x,y\}$, we have $z \pref{P} x$ if and only if $z \pref{P} y$. Given a preference ordering $P$ and a candidate $y \in \C$, let $P^y$ be the preference ordering $P$ except that $y$ is swapped with the candidate directly above $y$, if such a candidate exists. 

A voting rule $v: \P^n \to \Delta(\C)$ is said to be \emph{pairwise responsive} if for every preference profile $\vec{P} = (P_1, \ldots, P_n) \in \P^n$, every $i \in [n]$, and every pair of candidates $x,y \in \C$ such that $x \pref{P_i}! y$, we have $v(z,(\vec{P}_{-i},{P_i}^y)) = v(z,\vec{P})$ for every candidate $z \in \C \setminus \{x,y\}$. Intuitively, a voting rule is pairwise responsive if for any preference profile, if we take a voter's preference ordering and swap two adjacent candidates, then only the selection probabilities of the swapped candidates can possibly change; the selection probabilities of the other candidates (not involved in the swap) remain the same. This implies that if a voting rule is pairwise responsive and we swap two adjacent candidates in a voter's preference ordering, then the change in the selection probability of one of the swapped candidates is the exact opposite (i.e., the additive inverse) of the change in the selection probability of the other swapped candidate; this is because the selection probabilities of all the candidates must sum up to 1. 

A voting rule $v: \P^n \to \Delta(\C)$ is said to be \emph{pairwise isolated} if for every $x, y \in \C$, every $i \in [n]$, and every $\vec{P} = (P_1, \ldots, P_n), \vec{P'} = (P'_1, \ldots, P'_n) \in \P^n$ such that $x \pref{P_i}! y$, $P'_i = P_i$, and the relative ordering of $x$ and $y$ for $P_j$ is the same as that for $P'_j$ for every $j \in [n]$, we have $v(y,(\vec{P'}_{-i},{P'_i}^y)) - v(y,\vec{P'}) = v(y,(\vec{P}_{-i},{P_i}^y)) - v(y,\vec{P})$. Intuitively, a voting rule is pairwise isolated if for any preference profile, if we take a voter $i$'s preference ordering and swap two adjacent candidates $x$ and $y$ with $x$ being on top of $y$, then the change in the selection probability of $y$ only depends on voter $i$'s preference ordering and the relative ordering of $x$ and $y$ in the other voters' preference orderings.

It is already known that anonymous randomized weakly strategy-proof voting rules are both pairwise responsive and pairwise isolated (see \cite{McL11}). 

\begin{lemma}[\cite{McL11}]
\label{lem:basicProperties}
Let $v: \P^n \to \Delta(\C)$ be any anonymous randomized voting rule that is weakly strategy-proof. Then, $v$ is pairwise responsive and pairwise isolated. 
\end{lemma}

\begin{proof}
This immediately follows from Lemmas 1, 2, and 3 in \cite{McL11}, where Lemmas 2 and 3 in \cite{McL11} follow immediately from Lemmas 1 and 3 in \cite{Gib77}.
\end{proof}

We will use Lemma \ref{lem:basicProperties} at various places below. We now show that if a voting rule is weakly strategy-proof and satisfies $\eps$-super-weak unanimity, then it also satisfies $\eps$-strong unanimity.

\begin{lemma}
\label{lem:superWeakToStrong}
Let $v: \P^n \to \Delta(\C)$ be any anonymous randomized voting rule that is weakly strategy-proof and satisfies $\eps$-super-weak unanimity. Then, $v$ satisfies $\eps$-strong unanimity.
\end{lemma}

The proof of Lemma \ref{lem:superWeakToStrong} uses the pairwise responsive property of $v$ (Lemma \ref{lem:basicProperties}) to obtain $\eps$-strong unanimity from $\eps$-super-weak unanimity; see \ref{app:impossibility} for the proof of Lemma \ref{lem:superWeakToStrong}. Given a preference profile $\vec{P} = (P_1, \ldots, P_n) \in \P^n$, let $top(\vec{P}) = (top(P_1), \ldots, top(P_n))$. We now show that if a voting rule is weakly strategy-proof and satisfies $\eps$-super-weak unanimity, then the voting rule is ``close to'' being ``tops-only'', i.e., depending only on the vector $top(\vec{P})$ of top choices of the preference orderings in the input $\vec{P}$; the ordering of the candidates below the top choices only affect the probabilities of the voting rule by a small amount $O(\eps)$. 

\begin{lemma}
\label{lem:closeToTopsOnly}
Let $v: \P^n \to \Delta(\C)$ be any anonymous randomized voting rule that is weakly strategy-proof and satisfies $\eps$-super-weak unanimity. Then, for every pair of preference profiles $\vec{P}, \vec{P'} \in \P^n$ such that $top(\vec{P}) = top(\vec{P'})$, we have $|v(x,\vec{P}) - v(x,\vec{P'})| \leq m\eps$ for every $x \in \C$.
\end{lemma}

The proof of Lemma \ref{lem:closeToTopsOnly} uses the pairwise isolation property (Lemma \ref{lem:basicProperties}) and the $\eps$-strong unanimity property (Lemma \ref{lem:superWeakToStrong}); the greatest difficulty in the proof lies in ensuring that the error (the $m\eps$ in Lemma \ref{lem:closeToTopsOnly}) is $O(\eps)$ instead of $O(\eps n)$, which is much too large. See \ref{app:impossibility} for the proof of Lemma \ref{lem:closeToTopsOnly}. We now show that if a voting rule is weakly strategy-proof and satisfies $\eps$-super-weak unanimity, then the selection probability of any candidate $x$ is ``close to'' depending only on the number of voters with $x$ as the top choice; the top choices of the other voters only affect the selection probability of $x$ by a small amount $O(\eps)$. 

\begin{lemma}
\label{lem:timesAtTop}
Let $v: \P^n \to \Delta(\C)$ be any anonymous randomized voting rule that is weakly strategy-proof and satisfies $\eps$-super-weak unanimity. Then, for every candidate $x \in \C$, and every pair of preference profiles $\vec{P} = (P_1, \ldots, P_n), \vec{P'} = (P'_1, \ldots, P'_n) \in \P^n$ such that $|\{i \in [n] : top(P_i) = x\}| = |\{i \in [n] : top(P'_i) = x\}|$, we have $|v(x,\vec{P}) - v(x,\vec{P'})| \leq 2m\eps$.
\end{lemma}

See \ref{app:impossibility} for the proof of Lemma \ref{lem:timesAtTop}. 

We now use the above lemmas to prove Theorem \ref{thm:impossibility}. Suppose $|\C| \geq 3$. Fix $v: \P^n \to \Delta(\C)$ to be any anonymous randomized voting rule that is weakly strategy-proof and satisfies $\eps$-super-weak unanimity. By Lemma \ref{lem:basicProperties}, $v$ is pairwise responsive and pairwise isolated, and by Lemma \ref{lem:superWeakToStrong}, $v$ also satisfies $\eps$-strong unanimity. We know from Lemma \ref{lem:timesAtTop} that the selection probability of any candidate $x$ is close to depending only on the number of voters with $x$ as the top choice. Thus, we will define a function $v': \C \times \{0, \ldots, n\}$ that, on input a candidate $x \in \C$ and a number $j \in \{0, \ldots, n\}$, specifies the approximate selection probability of $x$ when exactly $j$ voters have $x$ as their top choice. 

Let $\C = \{a_1, \ldots, a_m\}$, where $a_1, \ldots, a_m$ is any fixed ordering of the candidates in $\C$. Let $v': \C \times \{0, \ldots, n\} \to [0,1]$ be defined by $v'(x,j) = v(x,\vec{P}^{x,j})$, where $\vec{P}^{x,j} = (P_1, \ldots, P_n)$ is the preference profile defined as follows: for $i = 1, \ldots, j$, let the top choice of $P_i$ be candidate $x$ and let the other candidates be ordered according to the ordering $a_1, \ldots, a_m$; for $i = j+1, \ldots n$, let $P_i$ be the ordering $a_1, \ldots, a_m$ with candidate $x$ moved to the bottom. By Lemma \ref{lem:timesAtTop}, the following claim follows immediately.

\begin{claim}
\label{claim:vCloseTov'}
For every preference profile $\vec{P} = (P_1, \ldots, P_n)$ with $j := |\{i \in [n] : top(P_i) = x\}|$, and every candidate $x \in \C$, we have $|v(x,\vec{P}) - v'(x,j)| \leq 2m\eps$.
\end{claim}

We now show that the candidates are ``close to being anonymous'' in the sense that changing the candidate in the input for $v'$ only changes the value of $v'$ by a small amount $O(\eps)$.

\begin{claim}
\label{claim:candidatesAreAnonymous}
For every pair of candidates $x,y \in \C$ and every $j \in \{0, \ldots, n\}$, we have $|v'(x,j) - v'(y,j)| \leq 14m\eps$.
\end{claim}

See \ref{app:impossibility} for the proof of Claim \ref{claim:candidatesAreAnonymous}. We now show that for any candidate $x \in \C$, the function $v'(x,\cdot)$ is ``close to being linear'' in the following sense: When the number of top choice votes for candidate $x$ (i.e., the $j$ in $v'(x,j)$) is increased by $\ell$, the change in $v'(x,\cdot)$ depends very little on how many top choice votes candidate $x$ had initially, which can only affect the change in $v'(x,\cdot)$ by a small amount $O(\eps)$. This roughly means that as the number of top choice votes for candidate $x$ increases, the selection probability of $x$ increases roughly linearly. 

\begin{claim}
\label{claim:slidingWindow}
Let $x \in \C$, let $j,j' \in \{0, \ldots, n-1\}$, and let $\ell \in [n]$ such that $j + \ell \leq n$ and $j' + \ell \leq n$. Then,
\begin{align*}
v'(x,j+\ell) - v'(x,j) = v'(x,j'+\ell) - v'(x,j') + \delta
\end{align*}
for some $\delta \in \R$ such that $|\delta| \leq 64m\eps$.
\end{claim}

See \ref{app:impossibility} for the proof of Claim \ref{claim:slidingWindow}. Using Claim \ref{claim:slidingWindow}, we now show that $v'(x,j)$ (which approximates the selection probability of candidate $x$ with $j$ top choice votes) is close to $\frac{j}{n}$, which is the selection probability of $x$ for the random dictatorship voting rule. Even though Claim \ref{claim:slidingWindow} says that $v'(x,j)$ is close to being linear, naive usage of Claim \ref{claim:slidingWindow} would result in $O(\eps n)$ error, which is too high; however, by using a better approach, we can make the error only $O(\eps)$.

\begin{claim}
\label{claim:closeToRandomDictatorship}
Let $x \in \C$. For every $j \in \{0, \ldots, n\}$, we have $|v'(x,j) - \frac{j}{n}| \leq O(\eps)$. 
\end{claim}

See \ref{app:impossibility} for the proof of Claim \ref{claim:closeToRandomDictatorship}.

We are now ready to complete the proof of Theorem \ref{thm:impossibility}, i.e., we will show that $v$ is $O(\eps)$-close to the random dictatorship voting rule. Let $\vec{P} = (P_1, \ldots, P_n) \in \P^n$, let $x \in \C$, and let $j = |\{i \in [n] : top(P_i) = x\}|$. We will show that $|v(x,\vec{P}) - \frac{j}{n}| \leq O(\eps)$. By Claim \ref{claim:vCloseTov'}, we have $|v(x,\vec{P}) - v'(x,j)| \leq O(\eps)$, and by Claim \ref{claim:closeToRandomDictatorship}, we also have $|v'(x,j) - \frac{j}{n}| \leq O(\eps)$. Thus, by the triangle inequality, we have $|v(x,\vec{P}) - \frac{j}{n}| \leq O(\eps)$, as required. This completes the proof of Theorem \ref{thm:impossibility}.

\section{Acknowledgments}

We thank Ron Rivest for helpful discussions. We also thank anonymous reviewers for helpful suggestions and comments.

\bibliographystyle{amsalpha}
\bibliography{VotingWithCoarseBeliefs}

\appendix
\renewcommand\thesection{Appendix \Alph{section}}

\section{Background Information on the Gibbard-Satterthwaite Theorem}
\label{app:Gibbard-Satterthwaite}

Roughly speaking, a voting rule is said to be \emph{strategy-proof} if no voter can gain utility in expectation by lying about her true preferences. We now give the formal definition of strategy-proof.

\begin{definition}[Strategy-proof]
A voting rule $v: \P^n \to \Delta(\C)$ is \emph{strategy-proof} if for every $i \in [n]$, every preference profile $\vec{P}_{-i} \in \P^{n-1}$, every pair of preference orderings $P_i, P_i' \in \P$, and every utility function $u_i$ that is consistent with $P_i$, we have
\begin{align*}
  \E[u_i(v(\vec{P}_{-i},P_i))] \geq \E[u_i(v(\vec{P}_{-i},P_i'))].
\end{align*}
\end{definition} 

It is desirable for a voting rule to be strategy-proof, since we can then expect voters to honestly submit their true preferences, and thus the candidate chosen by the voting rule will better reflect the voters' true preferences. Unfortunately, if there are at least three candidates, then it is not possible for a deterministic and onto voting rule to be strategy-proof unless it is \emph{dictatorial}, i.e., there exists some voter $i$ such that the voting rule simply always chooses voter $i$'s top choice. This was shown independently by Gibbard \cite{Gib73} and Satterthwaite \cite{Sat75}, and is known as the Gibbard-Satterthwaite theorem.

\begin{theorem}[Gibbard-Satterthwaite \cite{Gib73,Sat75}]
Suppose there are at least three candidates, i.e., $|\C| \geq 3$. Let $v: \P^n \to \C$ be any deterministic voting rule that is onto and strategy-proof. Then, $v$ is dictatorial, i.e., there exists an $i \in [n]$ such that $v(P_1, \ldots, P_n) = top(P_i)$ for every preference profile $(P_1, \ldots, P_n) \in \P^n$. 
\end{theorem}

The Gibbard-Satterthwaite theorem considers voting rules that are \emph{deterministic}. However, several years later, Gibbard \cite{Gib77} generalized the Gibbard-Satterthwaite theorem to \emph{randomized} voting rules. Before we state Gibbard's generalized impossibility result, let us state some required definitions. A (randomized) voting rule $v: \P^n \to \Delta(\C)$ is said to be \emph{unilateral} if it only depends on the preference of a single voter, i.e., there exists an $i \in [n]$ such that $v(\vec{P}) = v(\vec{P'})$ for every $\vec{P} = (P_1, \ldots, P_n), \vec{P'} = (P_1', \ldots, P_n') \in \P^n$ such that $P_i = P_i'$. A (randomized) voting rule $v: \P^n \to \Delta(\C)$ is said to be \emph{duple} if $v$ always chooses some candidate from a fixed set of two candidates, i.e., there exist candidates $x,y \in \C$ such that $v(z,\vec{P}) = 0$ for every $z \in \C \setminus \{x,y\}$ and $\vec{P} \in \P^n$. 

Intuitively, when there are at least three candidates, both unilateral rules and duple rules are undesirable, since the former only consider a single voter's preference, and the latter essentially ignore all but two candidates. Gibbard's generalized impossibility result \cite{Gib77} states that any randomized strategy-proof voting rule is a probability distribution over unilateral rules and duple rules.

\begin{theorem}[Gibbard \cite{Gib77}]
Let $v: \P^n \to \Delta(\C)$ be any randomized voting rule that is strategy-proof. Then, $v$ is a distribution over unilateral rules and duple rules, i.e., there exist randomized voting rules $v_1, \ldots, v_t$ and weights $\alpha_1, \ldots, \alpha_t \in (0,1]$ with $\sum_{i=1}^t \alpha_i = 1$, such that each $v_i$ is unilateral or duple, and $v(x,\vec{P}) = \alpha_1 v_1(x,\vec{P}) + \cdots + \alpha_t v_t(x,\vec{P})$ for every $\vec{P} \in \P^n$ and $x \in \C$.
\end{theorem}

A corollary of Gibbard's impossibility result is that if a randomized voting rule is strategy-proof and Pareto efficient, then it is a probability distribution over dictatorial voting rules.

\begin{corollary}[Gibbard \cite{Gib77}]
Let $v: \P^n \to \Delta(\C)$ be any randomized voting rule that is strategy-proof and Pareto efficient. Then, $v$ is a distribution over dictatorial voting rules, i.e., there exist dictatorial voting rules $v_1, \ldots, v_t$ and weights $\alpha_1, \ldots, \alpha_t \in (0,1]$ with $\sum_{i=1}^t \alpha_i = 1$, such that $v(x,\vec{P}) = \alpha_1 v_1(x,\vec{P}) + \cdots + \alpha_t v_t(x,\vec{P})$ for every $\vec{P} \in \P^n$ and $x \in \C$.
\end{corollary}

\section{Forming a Belief from Observations}
\label{app:formingBeliefs}

We now describe how forming a belief from observations using empirical frequencies or a Dirichlet distribution yields $\alpha$-coarse beliefs, where $\alpha$ is inversely proportional to the number of observations.

\paragraph{Forming a belief using empirical frequencies.} Consider a voter that forms a belief $\phi$ based on $\ell$ observations $P_1, \ldots, P_{\ell}$ by simply setting $\phi(P)$ to be the relative frequency of $P$ in $P_1, \ldots, P_{\ell}$, i.e., $\phi(P) = \frac{|\{j \in [\ell] : P_j = P\}|}{\ell}$. We see that the resulting belief $\phi$ is $(1/\ell)$-coarse. 

\paragraph{Forming a belief using a Dirichlet distribution.} Let us first describe a common method of forming a belief based on observations of preferences. We begin with some initial distribution (e.g., the uniform distribution) over the set of all beliefs, and as we make observations, we update this distribution using Bayes' Rule. At any time, our distribution over beliefs can be used to form a single belief by taking the expectation of the distribution over beliefs; equivalently, the single belief is the resulting distribution over preferences obtained by first sampling a belief from the distribution over beliefs, and then sampling a preference from the sampled belief. 

At the beginning when no samples of preferences have been observed yet, we are indifferent between different possible beliefs, so we start with the uniform distribution over the set $\Delta(\P)$ of all beliefs. Then, given an observation of a preference ordering $P_1$, we update the uniform distribution over $\Delta(\P)$ by conditioning on the event that the sample $P_1$ is observed. Upon further observations $P_2, \ldots, P_\ell$, we update the current distribution over $\Delta(\P)$ by conditioning on each of the observations $P_2, \ldots, P_\ell$ separately in sequence. The resulting distribution over beliefs can be ``collapsed'' to give us a single belief as described above.

The distributions over beliefs that we obtain can be described by the \emph{Dirichlet distribution}. The Dirichlet distribution $Dir(\vec{\alpha})$ of order $K \geq 2$ with parameters $\vec{\alpha} = (\alpha_1, \ldots, \alpha_K) > 0$ has a pdf given by $f_{\vec{\alpha}}(x_1, \ldots, x_K) \sim \prod_{i=1}^K x_i^{\alpha_i - 1}$ for every $(x_1, \ldots, x_K) \in \R^K$ such that $\sum_{i=1}^K x_i = 1$, and is 0 elsewhere. In our context of updating beliefs, we fix an arbitrary ordering of the preferences in $\P$, and we let $K = |\P|$, so $(x_1, \ldots, x_K)$ (with $\sum_{i=1}^K x_i = 1$) are the probability masses describing a belief. The uniform distribution over the set of all beliefs is the Dirichlet distribution $Dir(1, \ldots, 1)$. It is known that if the current distribution over beliefs is $Dir(\vec{\alpha})$ and we observe $P_1, \ldots, P_\ell$, then the resulting distribution over beliefs obtained by conditioning on $P_1, \ldots, P_\ell$ is $Dir(\vec{\alpha} + \vec{c})$, where $\vec{c}$ is the vector of counts representing how many times each preference ordering appears in the observations $P_1, \ldots, P_\ell$. It is also known that for any $\vec{\alpha'} = (\alpha'_1, \ldots, \alpha'_K)$, the expectation of $Dir(\vec{\alpha'})$ is $\frac{1}{\sum_{i=1}^K \alpha'_i} \cdot (\alpha'_1, \ldots, \alpha'_K)$. Let $\vec{\alpha} = (1, \ldots, 1)$ (vector of $K$ 1's), and let $\vec{\alpha'} = \vec{\alpha} + \vec{c}$, where $\vec{c}$ is as described above. Noting that $||\vec{c}||_1 = \ell$ (since there are $\ell$ observations), the expectation of $Dir(\vec{\alpha'})$ is $\frac{1}{K+\ell} \cdot (\alpha'_1, \ldots, \alpha'_K)$. Since the belief formed from $Dir(\vec{\alpha'})$ is the expectation of $Dir(\vec{\alpha'})$, and since the $\alpha'_i$'s are integers, we see that the obtained belief is $\frac{1}{K+\ell}$-coarse.

\section{Proofs for Section \ref{sec:framework} and \ref{sec:examples}}
\label{app:ProofsFromFrameworkAndExamples}

\begin{reptheorem}{thm:framework}
Let $f: \P^* \to \Delta(2^{\C})$ be any elimination rule, and let $s = \{s_A\}_{A \subseteq \C, A \neq \emptyset}$ be any selection rule. Let $\delta : \mathbb{N} \to \R$ be any function. Suppose there exists a polynomial $p(\cdot)$ such that for every $\alpha > 0$ and every $n \geq p(\frac{1}{\alpha})$, the following holds: 
\begin{itemize} 
	\item For every $i \in [n]$ and every $\alpha$-coarse i.i.d.~belief $\phi_i$, there exists a non-empty subset $A \subseteq \C$ of candidates such that the following conditions hold:
	\begin{itemize}
		\item For every $P_i \in \P$, the elimination rule $f(\vec{P}_{-i}, P_i)$ chooses (to keep) $A$ with probability at least $1 - \delta(n)$ over the randomness of $\vec{P}_{-i} \sim {\phi_i}^{n-1}$ and $f$.
		\item $s_A$ is strategy-proof w.r.t.~the restricted belief $\phi_i|_A$. 
	\end{itemize}
\end{itemize}
Then, the voting rule $v: \P^* \to \Delta(\C)$ defined by $v(\vec{P}) = s(\vec{P}|_{f(\vec{P})})$ is large-scale $2\delta$-strategy-proof w.r.t.~coarse i.i.d.~beliefs, and the rate of $v$ is the polynomial $p(\cdot)$.
\end{reptheorem}

\begin{proof}
Let $\alpha > 0$, let $n \geq p(1/\alpha)$, let $i \in [n]$, let $P_i, P_i' \in \P$, let $\phi_i$ be any $\alpha$-coarse i.i.d.~belief, and let $u_i$ be any utility function that is consistent with $P_i$. Let $\vec{P}_{-i} \sim {\phi_i}^{n-1}$. We will show that 
\begin{align*}
  \E[u_i(v(\vec{P}_{-i},P_i))] \geq \E[u_i(v(\vec{P}_{-i},P_i'))] - 2\delta(n). \tag{1}
\end{align*}
Let $A$ be the set of candidates guaranteed by the assumptions of the theorem statement. Consider an alternate voting rule $v': \P^* \to \Delta(\C)$ defined by $v'(\vec{P}) = s(\vec{P}|_A)$. Since the elimination rule $f(\vec{P}_{-i}, P_i)$ chooses $A$ with probability at least $1 - \delta(n)$, it is easy to see that for every $P \in \P$, we have $||v(\vec{P}_{-i},P) - v'(\vec{P}_{-i},P)||_1 \leq \delta(n)$. Thus, we have
\begin{align*}
& |\E[u_i(v(\vec{P}_{-i},P_i))] - \E[u_i(v'(\vec{P}_{-i},P_i))]| \\
= \ & \left|\sum_{c \in \C} \Pr[v(\vec{P}_{-i},P_i) = c] \cdot u_i(c) - \sum_{c \in \C} \Pr[v'(\vec{P}_{-i},P_i) = c] \cdot u_i(c)\right| \\
\leq \ & \sum_{c \in \C} |\Pr[v(\vec{P}_{-i},P_i) = c] - \Pr[v'(\vec{P}_{-i},P_i) = c]| \cdot |u_i(c)| \\
\leq \ & \delta(n). \tag{2}
\end{align*} 
Similarly, we also have
\begin{align*}
|\E[u_i(v(\vec{P}_{-i},P_i'))] - \E[u_i(v'(\vec{P}_{-i},P_i'))]| \leq \delta(n). \tag{3}
\end{align*} 
Since $s_A$ is strategy-proof w.r.t.~$\phi_i|_A$, we have $\E[u_i(v'(\vec{P}_{-i},P_i))] \geq \E[u_i(v'(\vec{P}_{-i},P_i'))]$. Combining this with (2) and (3) yields (1), as required.
\end{proof}

\begin{reptheorem}{thm:v_pl}
Let $0 < \delta < 1/2$, and let $v_{pl}$ be the voting rule defined above Theorem \ref{thm:v_pl} in the body of the paper. Then, $v_{pl}$ satisfies the following properties:
\begin{enumerate}
	\item $v_{pl}$ is large-scale $(e^{-\Omega(n^{2\delta})})$-strategy-proof w.r.t.~coarse i.i.d.~beliefs, with rate $p(x) = O(x^{\lceil 1/(1/2-\delta) \rceil})$.
	\item $v_{pl}$ is Pareto efficient.
	\item $v_{pl}$ is $n^{1/2+\delta}$-close to optimal in the sense that $v_{pl}$ always chooses a candidate $c \in \C$ such that the number of top-choice votes for $c$ is within $n^{1/2+\delta}$ of the highest number of top-choice votes among the candidates.
\end{enumerate}
\end{reptheorem}

\begin{proof}
Property 3 clearly follows from the definition of $v_{pl}$. We will now show Property 2. Let $\vec{P} \in \P^n$ be a preference profile such that every voter in $\vec{P}$ prefers candidate $x$ over candidate $y$. We note that in order for candidate $y$ to be chosen as the winner, candidate $y$ must be in the set of remaining candidates in Stage 2. However, when this occurs, candidate $x$ would also be in the set of remaining candidates in Stage 2, since candidate $x$ always has a count that is higher than that of candidate $y$. Thus, candidate $y$ would have no top-choice votes in Stage 2, so it cannot be chosen as the winner by the plurality rule in Stage 2. We have now shown Property 2. 

We will now show Property 1. We will use our general framework, i.e., Theorem \ref{thm:framework}. The elimination rule $f: \P^* \to \Delta(2^{\C})$ corresponds to Stage 1, i.e., it chooses to keep the candidates that are remaining at the end of Stage 1. The selection rule $s = \{s_A\}_{A \subseteq \C, A \neq \emptyset}$ runs the plurality rule on the remaining candidates with respect to the restricted preference profile. For each non-empty $A \subseteq \C$, let $\Phi_A'$ be the set of beliefs $\phi$ (over the set of all preference orderings on $A$) where every candidate in $A$ has the same probability of being the top choice. It is not hard to verify that the plurality rule, and thus the selection rule, is strategy-proof with respect to each $\Phi_A'$. Let $p(x) = (3x+1)^{\lceil 1/(1/2-\delta) \rceil}$, let $\alpha > 0$, let $n \geq p(1/\alpha)$, let $i \in [n]$, and let $\phi_i$ be any $\alpha$-coarse i.i.d.~belief. 

Given a preference ordering $P$ and a candidate $x \in \C$, let $points(x,P)$ be 1 if $x$ is the top choice in $P$, and 0 otherwise. Let $A$ be the set of candidates remaining after the following procedure:
\begin{itemize}
	\item Let $S = \C$, and repeatedly do the following until no more candidates are eliminated: Eliminate all the candidates $a \in S$ such that $\E_{P \sim \phi_i}[points(a,P|_S)] < \max_{a' \in S} \E_{P \sim \phi_i}[points(a',P|_S)]$, and let $S$ be the set of remaining candidates.
\end{itemize}
We will show that the following conditions hold:
\begin{itemize} 
	\item For every $P_i \in \P$, the elimination rule $f(\vec{P}_{-i}, P_i)$ chooses (to keep) $A$ with probability at least $1 - e^{-\Omega(n^{2\delta})}$ over the randomness of $\vec{P}_{-i} \sim {\phi_i}^{n-1}$ and $f$.
	\item The restriction of $\phi_i$ to $A$ results in a belief in $\Phi'_A$. 
\end{itemize}
The second condition holds because from the definition of $A$, we see that for every $a \in A$, we have $\Pr_{P \sim \phi_i}[top(P|_A)=a] = \E_{P \sim \phi_i}[points(a,P|_A)] = \max_{a' \in A} \E_{P \sim \phi_i}[points(a',P|_A)]$. Thus, we now show the first condition. 

Let $P_i \in \P$. Consider the execution of one iteration of the loop in Stage 1. Suppose the current set of remaining candidates is $S$ and we are currently at Stage 1. Let $M = \max_{a' \in S} \E_{P \sim \phi_i}[points(a',P|_S)]$. Let $E$ be the set of candidates $a \in S$ such that $\E_{P \sim \phi_i}[points(a,P|_S)] = M$. We first show that for each candidate $y \in S \setminus E$, we have 
\begin{align*}
\E_{P \sim \phi_i}[points(y,P|_S)] \leq M - \alpha. \tag{1}
\end{align*}
It suffices to show that for every $x,y \in S$, we have $|\E_{P \sim \phi_i}[points(x,P|_S)] - \E_{P \sim \phi_i}[points(y,P|_S)]| = 0$ or $|\E_{P \sim \phi_i}[points(x,P|_S)] - \E_{P \sim \phi_i}[points(y,P|_S)]| \geq \alpha$. To see this, let $x,y \in S$, and observe that
\begin{align*}
\ & |\E_{P \sim \phi_i}[points(x,P|_S)] - \E_{P \sim \phi_i}[points(y,P|_S)]| \\
= \ & \left|\sum_{P \in \P} \phi_i(P) \cdot points(x,P|_S) - \sum_{P \in \P} \phi_i(P) \cdot points(y,P|_S)\right| \\
= \ & \left|\sum_{P \in \P} \phi_i(P) \cdot (points(x,P|_S) - points(y,P|_S))\right|. \tag{2}
\end{align*}
Since $\phi_i$ is $\alpha$-coarse, there exists a $\beta \geq \alpha$ such that for every $P \in \P$, $\phi_i(P)$ is a multiple of $\beta$. Thus, each term of the sum in (2) is a multiple of $\beta$, and so the sum is also a multiple of $\beta$. Thus, the entire expression in (2) is either 0 or at least $\beta \geq \alpha$, as required. Thus, we have shown (1).

Let $P_{i'} \sim \phi_i$ independently for every $i' \in [n] \setminus \{i\}$, and let $score_{-i}(x) = \sum_{i' \in [n] \setminus \{i\}} points(x,P_{i'}|_S)$ for every $x \in S$. 
By a Chernoff bound, for each $x \in S$, we have 
\begin{align*}
\Pr[|score_{-i}(x) - \E[score_{-i}(x)]| \geq n^{1/2 + \delta} / 4] \leq e^{-\Omega(n^{2\delta})}. 
\end{align*}
Now, by the union bound, we have
\begin{align*}
\Pr[\exists x \in S : |score_{-i}(x) - \E[score_{-i}(x)]| \geq n^{1/2 + \delta} / 4] \leq m \cdot e^{-\Omega(n^{2\delta})} = e^{-\Omega(n^{2\delta})}. \tag{3}
\end{align*}
Since $\E[score_{-i}(x)] = (n-1) \cdot M$ for every $x \in E$, it follows from (3) that 
\begin{align*}
\Pr[\exists x,y \in E : |score_{-i}(x) - score_{-i}(y)| \geq n^{1/2 + \delta} / 2] \leq e^{-\Omega(n^{2\delta})}. \tag{4}
\end{align*}
From (1), we have $\E_{P \sim \phi_i}[points(y,P|_S)] \leq M - \alpha$ for every $y \in S \setminus E$. Thus, $\E[score_{-i}(y)] = (n-1) \cdot \E_{P \sim \phi_i}[points(y,P|_S)] \leq (n-1) M - (n-1) \alpha < (n-1) M - 2n^{1/2 + \delta}$ for every $y \in S \setminus E$, so it also follows from (3) that 
\begin{align*}
\Pr[\exists x \in S \setminus E, y \in E : score_{-i}(x) \geq score_{-i}(y) - n^{1/2 + \delta} - 1] \leq e^{-\Omega(n^{2\delta})}. \tag{5}
\end{align*}
Since the elimination rule $f$ eliminates precisely the candidates that have a score (i.e., count) that is not within $n^{1/2 + \delta}$ of the maximum score among the candidates, and since voter $i$'s preference ordering $P_i$ adds at most $1$ to the score of a candidate, we see (from (4), (5), and the union bound) that with probability at least $1 - e^{-\Omega(n^{2\delta})}$, precisely the candidates in $S \setminus E$ will be eliminated in the current iteration of Stage 1. Thus, at each iteration, with probability at least $1 - e^{-\Omega(n^{2\delta})}$, the set of candidates that get eliminated in the iteration precisely matches the set of candidates that would be eliminated in the procedure used to define $A$. Thus, by the union bound, we have that with probability at least $1 - m \cdot e^{-\Omega(n^{2\delta})} = 1 - e^{-\Omega(n^{2\delta})}$, the elimination rule $f(\vec{P}_{-i}, P_i)$ chooses (to keep) $A$.

Now, by Theorem \ref{thm:framework}, $v_{pl}$ is large-scale $e^{-\Omega(n^{2\delta})}$-strategy-proof w.r.t.~coarse i.i.d.~beliefs, with rate $p(x) = O(x^{\lceil 1/(1/2-\delta) \rceil})$.
\end{proof}

\begin{reptheorem}{thm:v_irv}
Let $0 < \delta < 1/2$, and let $v_{irv}$ be the voting rule defined above Theorem \ref{thm:v_irv} in the body of the paper. Then, $v_{irv}$ satisfies the following properties:
\begin{enumerate}
	\item $v_{irv}$ is large-scale $(e^{-\Omega(n^{2\delta})})$-strategy-proof w.r.t.~coarse i.i.d.~beliefs, with rate $p(x) = O(x^{\lceil 1/(1/2-\delta) \rceil})$.
	\item $v_{irv}$ is Pareto efficient.
\end{enumerate}
\end{reptheorem}

\begin{proof}
We first show Property 2. Let $\vec{P} \in \P^n$ be a preference profile such that every voter in $\vec{P}$ prefers candidate $x$ over candidate $y$. We note that in order for candidate $y$ to be chosen as the winner, candidate $y$ must be in the set of remaining candidates in Stage 2. However, when this occurs, candidate $x$ would also be in the set of remaining candidates in Stage 2, since candidate $x$ always has a count that is higher than that of candidate $y$. Thus, candidate $y$ would have no top-choice votes in Stage 2, so it cannot be chosen as the winner by the plurality rule in Stage 2. We have now shown Property 2. 

We will now show Property 1. We will use our general framework, i.e., Theorem \ref{thm:framework}. The elimination rule $f: \P^* \to \Delta(2^{\C})$ corresponds to Stage 1, i.e., it chooses to keep the candidates that are remaining at the end of Stage 1. The selection rule $s = \{s_A\}_{A \subseteq \C, A \neq \emptyset}$ runs the plurality rule on the remaining candidates with respect to the restricted preference profile. For each non-empty $A \subseteq \C$, let $\Phi_A'$ be the set of beliefs $\phi$ (over the set of all preference orderings on $A$) where every candidate in $A$ has the same probability of being the top choice. It is not hard to verify that the plurality rule, and thus the selection rule, is strategy-proof with respect to each $\Phi_A'$. Let $p(x) = (3x+1)^{\lceil 1/(1/2-\delta) \rceil}$, let $\alpha > 0$, let $n \geq p(1/\alpha)$, let $i \in [n]$, and let $\phi_i$ be any $\alpha$-coarse i.i.d.~belief. 

Given a preference ordering $P$ and a candidate $x \in \C$, let $points(x,P)$ be 1 if $x$ is the top choice in $P$, and 0 otherwise. Let $A$ be the set of candidates remaining after the following procedure:
\begin{itemize}
	\item Initialize $S := \C$, and repeat the following: Eliminate all the candidates $a \in S$ such that $\E_{P \sim \phi_i}[points(a,P|_S)] = \min_{a' \in S} \E_{P \sim \phi_i}[points(a',P|_S)]$, unless this would eliminate all the remaining candidates, in which case we stop and exit the repeat loop without eliminating any of the remaining candidates. Let $S$ be the new set of remaining candidates.
\end{itemize}
We will show that the following conditions hold:
\begin{itemize} 
	\item For every $P_i \in \P$, the elimination rule $f(\vec{P}_{-i}, P_i)$ chooses (to keep) $A$ with probability at least $1 - e^{-\Omega(n^{2\delta})}$ over the randomness of $\vec{P}_{-i} \sim {\phi_i}^{n-1}$ and $f$.
	\item The restriction of $\phi_i$ to $A$ results in a belief in $\Phi'_A$. 
\end{itemize}
The second condition holds because from the definition of $A$, we see that for every $a \in A$, we have $\Pr_{P \sim \phi_i}[top(P|_A)=a] = \E_{P \sim \phi_i}[points(a,P|_A)] = \max_{a' \in A} \E_{P \sim \phi_i}[points(a',P|_A)]$. Thus, we now show the first condition. 

Let $P_i \in \P$. Consider the execution of one iteration of the loop in Stage 1. Suppose the current set of remaining candidates is $S$ and we are currently at Stage 1. Let $M = \min_{a' \in S} \E_{P \sim \phi_i}[points(a',P|_S)]$. Let $E$ be the set of candidates $a \in S$ such that $\E_{P \sim \phi_i}[points(a,P|_S)] = M$. We first show that for each candidate $y \in S \setminus E$, we have 
\begin{align*}
\E_{P \sim \phi_i}[points(y,P|_S)] \geq M + \alpha. \tag{1}
\end{align*}
It suffices to show that for every $x,y \in S$, we have $|\E_{P \sim \phi_i}[points(x,P|_S)] - \E_{P \sim \phi_i}[points(y,P|_S)]| = 0$ or $|\E_{P \sim \phi_i}[points(x,P|_S)] - \E_{P \sim \phi_i}[points(y,P|_S)]| \geq \alpha$. To this end, let $x,y \in S$, and observe that
\begin{align*}
\ & |\E_{P \sim \phi_i}[points(x,P|_S)] - \E_{P \sim \phi_i}[points(y,P|_S)]| \\
= \ & \left|\sum_{P \in \P} \phi_i(P) \cdot points(x,P|_S) - \sum_{P \in \P} \phi_i(P) \cdot points(y,P|_S)\right| \\
= \ & \left|\sum_{P \in \P} \phi_i(P) \cdot (points(x,P|_S) - points(y,P|_S))\right|. \tag{2}
\end{align*}
Since $\phi_i$ is $\alpha$-coarse, there exists a $\beta \geq \alpha$ such that for every $P \in \P$, $\phi_i(P)$ is a multiple of $\beta$. Thus, each term of the sum in (2) is a multiple of $\beta$, and so the sum is also a multiple of $\beta$. Thus, the entire expression in (2) is either 0 or at least $\beta \geq \alpha$, as required. Thus, we have shown (1).

Let $P_{i'} \sim \phi_i$ independently for every $i' \in [n] \setminus \{i\}$, and let $score_{-i}(x) = \sum_{i' \in [n] \setminus \{i\}} points(x,P_{i'}|_S)$ for every $x \in S$. 
By a Chernoff bound, for each $x \in S$, we have 
\begin{align*}
\Pr[|score_{-i}(x) - \E[score_{-i}(x)]| \geq n^{1/2 + \delta} / 4] \leq e^{-\Omega(n^{2\delta})}. 
\end{align*}
Now, by the union bound, we have
\begin{align*}
\Pr[\exists x \in S : |score_{-i}(x) - \E[score_{-i}(x)]| \geq n^{1/2 + \delta} / 4] \leq m \cdot e^{-\Omega(n^{2\delta})} = e^{-\Omega(n^{2\delta})}. \tag{3}
\end{align*}
Since $\E[score_{-i}(x)] = (n-1) \cdot M$ for every $x \in E$, it follows from (3) that 
\begin{align*}
\Pr[\exists x,y \in E : |score_{-i}(x) - score_{-i}(y)| \geq n^{1/2 + \delta} / 2] \leq e^{-\Omega(n^{2\delta})}. \tag{4}
\end{align*}
From (1), we have $\E_{P \sim \phi_i}[points(y,P|_S)] \geq M + \alpha$ for every $y \in S \setminus E$. Thus, $\E[score_{-i}(y)] = (n-1) \cdot \E_{P \sim \phi_i}[points(y,P|_S)] \geq (n-1) M + (n-1) \alpha > (n-1) M + 2n^{1/2 + \delta}$ for every $y \in S \setminus E$, so it also follows from (3) that 
\begin{align*}
\Pr[\exists x \in S \setminus E, y \in E : score_{-i}(x) \leq score_{-i}(y) + n^{1/2 + \delta} + 1] \leq e^{-\Omega(n^{2\delta})}. \tag{5}
\end{align*}
Since the elimination rule $f$ eliminates precisely the candidates that have a score (i.e., count) that is not within $n^{1/2 + \delta}$ of the minimum score among the candidates, and since voter $i$'s preference ordering $P_i$ adds at most $1$ to the score of a candidate, we see (from (4), (5), and the union bound) that with probability at least $1 - e^{-\Omega(n^{2\delta})}$, precisely the candidates in $E$ will be eliminated in the current iteration of Stage 1. Thus, at each iteration, with probability at least $1 - e^{-\Omega(n^{2\delta})}$, the set of candidates that get eliminated in the iteration precisely matches the set of candidates that would be eliminated in the procedure used to define $A$. Thus, by the union bound, we have that with probability at least $1 - m \cdot e^{-\Omega(n^{2\delta})} = 1 - e^{-\Omega(n^{2\delta})}$, the elimination rule $f(\vec{P}_{-i}, P_i)$ chooses (to keep) $A$.

Now, by Theorem \ref{thm:framework}, $v_{irv}$ is large-scale $e^{-\Omega(n^{2\delta})}$-strategy-proof w.r.t.~coarse i.i.d.~beliefs, with rate $p(x) = O(x^{\lceil 1/(1/2-\delta) \rceil})$.
\end{proof}

\section{More Examples of our General Framework}
\label{app:moreExamples}

A (positional) \emph{scoring rule} is a voting rule where each candidate $x$ receives a certain number of points from each voter $i$ depending on the position of $x$ in voter $i$'s preference ordering, and the candidate with the highest total score wins (breaking ties in some way). A scoring rule has a \emph{points vector} $(p_1, \ldots, p_m) \in \N^m$ associated with it; for each voter $i$ with submitted preference ordering $P_i$, the $j^{th}$ top candidate in $P_i$ receives $p_j$ points. There are many well-known examples of scoring rules, such as the following:
\begin{itemize}
  \item Plurality: The \emph{plurality} voting rule chooses the candidate with the most top-choice votes (breaking ties in some way). This is simply a scoring rule with the points vector $(1, 0, \ldots, 0) \in \N^m$. 
  \item Borda count: The \emph{Borda count} voting rule is a scoring rule with the points vector $(m, m-1, \ldots, 1) \in \N^m$ (recall that $m$ is the number of candidates).
\end{itemize}

\begin{example}[{\bf Scoring Rule Elimination + Input-Independent Selection}]
Let $0 < \delta < 1/2$, and let $v_{score}: \P^* \to \Delta(\C)$ be any voting rule defined as follows; on input a preference profile $\vec{P} \in \P^n$, $v_{score}$ does the following:
\begin{description}
	\item[Stage 1:] Use a scoring rule to compute the scores of the candidates, and then eliminate all the candidates with a score that is not within $n^{1/2+\delta}$ of the highest score among the candidates.
  \item[Stage 2:] Choose a winner (deterministically or randomly) from the remaining candidates in any way that does not depend on the input preference profile. 
\end{description}

Using our general framework, we now show that $v_{score}$ is large-scale $\eps$-strategy-proof w.r.t.~coarse i.i.d.~beliefs (where $\eps$ is exponentially small), and also satisfies certain efficiency properties.

\begin{theorem}
\label{thm:v_score}
Let $0 < \delta < 1/2$, and let $v_{score}$ be the voting rule defined above. Then, $v_{score}$ satisfies the following properties:
\begin{enumerate}
	\item $v_{score}$ is large-scale $(e^{-\Omega(n^{2\delta})})$-strategy-proof w.r.t.~coarse i.i.d.~beliefs, with rate $p(x) = O(x^{\lceil 1/(1/2-\delta) \rceil})$.
	\item $v_{score}$ is Pareto efficient if the points vector of the scoring rule is strictly decreasing, or if the scoring rule is the plurality rule and $n$ is sufficiently large.
	\item $v_{score}$ is $n^{1/2+\delta}$-close to optimal in the sense that $v_{score}$ always chooses a candidate $c \in \C$ such that the score of $c$ is within $n^{1/2+\delta}$ of the highest score among the candidates.
\end{enumerate}
\end{theorem}

\begin{proof}
Property 3 clearly follows from the definition of $v_{score}$. We will now show Property 2. Let $\vec{P} \in \P^n$ be a preference profile such that every voter in $\vec{P}$ prefers candidate $x$ over candidate $y$. It suffices to show that candidate $y$ will be eliminated by $v_{score}$, i.e., the score of $y$ is not within $n^{1/2 + \delta}$ of the maximum score among the candidates. If the points vector of the scoring rule is strictly decreasing, then the score of $x$ is at least $n$ more than the score of $y$ (since the points in the points vector are integers), as required. On the other hand, if the scoring rule is the plurality rule, then the score of $y$ is 0 while the maximum score among the candidates is at least $n / (|\C|-1)$; when $n$ is sufficiently large, the score of $y$ is not within $n^{1/2 + \delta}$ of the maximum score among the candidates, as required. We have shown Property 2. 

We will now show Property 1. We will use our general framework, i.e., Theorem \ref{thm:framework}. The elimination rule $f: \P^* \to \Delta(2^{\C})$ corresponds to Stage 1, i.e., $f$ chooses to keep the candidates that are within $n^{1/2+\delta}$ of the maximum score among the candidates. The selection rule $s = \{s_A\}_{A \subseteq \C, A \neq \emptyset}$ is the rule used in Stage 2. Clearly, for every non-empty $A \subseteq \C$, $s_A$ is strategy-proof with respect to the set of all beliefs. Let $p(x) = (3x+1)^{\lceil 1/(1/2-\delta) \rceil}$, let $\alpha > 0$, let $n \geq p(1/\alpha)$, let $i \in [n]$, and let $\phi_i$ be any $\alpha$-coarse i.i.d.~belief. 

Let $(p_1, \ldots, p_m) \in \N^m$ be the points vector associated with $v_{score}$. Given a preference ordering $P$ and a candidate $x \in \C$, let $points(x,P)$ be the number of points candidate $x$ would receive from a voter with submitted preference ordering $P$, i.e., $points(x,P) = p_j$, where $j$ is the position of candidate $x$ in $P$, with the topmost position being position 1.

Let $M = \max_{a' \in \C} \E_{P \sim \phi_i}[points(a',P)]$. Let $A$ be the set of candidates $a \in \C$ such that $\E_{P \sim \phi_i}[points(a,P)] = M$. We will show that the following holds:
\begin{itemize} 
	\item For every $P_i \in \P$, the elimination rule $f(\vec{P}_{-i}, P_i)$ chooses (to keep) $A$ with probability at least $1 - e^{-\Omega(n^{2\delta})}$ over the randomness of $\vec{P}_{-i} \sim {\phi_i}^{n-1}$ and $f$.
\end{itemize}
Let $P_i \in \P$. We first show that for each candidate $y \in \C \setminus A$, we have 
\begin{align*}
\E_{P \sim \phi_i}[points(y,P)] \leq M - \alpha. \tag{1}
\end{align*}
It suffices to show that for every $x,y \in \C$, we have $|\E_{P \sim \phi_i}[points(x,P)] - \E_{P \sim \phi_i}[points(y,P)]| = 0$ or $|\E_{P \sim \phi_i}[points(x,P)] - \E_{P \sim \phi_i}[points(y,P)]| \geq \alpha$. To see this, let $x,y \in \C$, and observe that
\begin{align*}
\ & |\E_{P \sim \phi_i}[points(x,P)] - \E_{P \sim \phi_i}[points(y,P)]| \\
= \ & \left|\sum_{P \in \P} \phi_i(P) \cdot points(x,P) - \sum_{P \in \P} \phi_i(P) \cdot points(y,P)\right| \\
= \ & \left|\sum_{P \in \P} \phi_i(P) \cdot (points(x,P) - points(y,P))\right|. \tag{2}
\end{align*}
Since $\phi_i$ is $\alpha$-coarse, there exists a $\beta \geq \alpha$ such that for every $P \in \P$, $\phi_i(P)$ is a multiple of $\beta$. Thus, each term of the sum in (2) is a multiple of $\beta$, and so the sum is also a multiple of $\beta$. Thus, the entire expression in (2) is either 0 or at least $\beta \geq \alpha$, as required. Thus, we have shown (1).

Let $P_{i'} \sim \phi_i$ independently for every $i' \in [n] \setminus \{i\}$, and let $score_{-i}(x) = \sum_{i' \in [n] \setminus \{i\}} points(x,P_{i'})$ for every $x \in \C$. By a Chernoff bound, for each $y \in \C$, we have 
\begin{align*}
\Pr[|score_{-i}(y) - \E[score_{-i}(y)]| \geq n^{1/2 + \delta} / 4] \leq e^{-\Omega(n^{2\delta})}. 
\end{align*}
Now, by the union bound, we have
\begin{align*}
\Pr[\exists x \in \C : |score_{-i}(x) - \E[score_{-i}(x)]| \geq n^{1/2 + \delta} / 4] \leq m \cdot e^{-\Omega(n^{2\delta})} = e^{-\Omega(n^{2\delta})}. \tag{3}
\end{align*}
Since $\E[score_{-i}(x)] = (n-1) \cdot M$ for every $x \in A$, it follows from (3) that 
\begin{align*}
\Pr[\exists x,y \in A : |score_{-i}(x) - score_{-i}(y)| \geq n^{1/2 + \delta} / 2] \leq e^{-\Omega(n^{2\delta})}. \tag{4}
\end{align*}
From (1), we have $\E_{P \sim \phi_i}[points(y,P)] \leq M - \alpha$ for every $y \in \C \setminus A$. Thus, $\E[score_{-i}(y)] = (n-1) \cdot \E_{P \sim \phi_i}[points(y,P)] \leq (n-1) M - (n-1) \alpha < (n-1) M - 2n^{1/2 + \delta}$ for every $y \in \C \setminus A$, so it also follows from (3) that 
\begin{align*}
\Pr[\exists x \in \C \setminus A, y \in A : score_{-i}(x) \geq score_{-i}(y) - n^{1/2 + \delta} - \max_{j \in [m]} p_j] \leq e^{-\Omega(n^{2\delta})}. \tag{5}
\end{align*}
Since the elimination rule $f$ eliminates precisely the candidates that have a score not within $n^{1/2 + \delta}$ of the maximum score among the candidates, and since voter $i$'s preference ordering $P_i$ adds at most $\max_{j \in [m]} p_j$ to the score of any candidate, we see (from (4), (5), and the union bound) that with probability at least $1 - e^{-\Omega(n^{2\delta})}$, the elimination rule $f(\vec{P}_{-i}, P_i)$ chooses to keep $A$.

Now, by Theorem \ref{thm:framework}, $v_{score}$ is large-scale $e^{-\Omega(n^{2\delta})}$-strategy-proof w.r.t.~coarse i.i.d.~beliefs, with rate $p(x) = O(x^{\lceil 1/(1/2-\delta) \rceil})$.
\end{proof}

\paragraph{Using plurality when there are only two candidates remaining.} Whenever Stage 1 eliminates all but two candidates, the voting rule $v_{score}$ can actually run the plurality rule on the two remaining candidates in Stage 2 instead of choosing a winner in a way that does not depend on the input preference profile. This is because the plurality rule is strategy-proof when there are only two candidates, and so the selection rule clearly still satisfies the requirements of our general framework. This improvement to the voting rule $v_{score}$ can be especially useful when it is widely believed that there are two ``strong'' candidates that are much more preferred by the voters than the other candidates. 
\end{example}

\section{Proofs for Section \ref{sec:punishingVotingRule}}
\label{app:punishingVotingRule}

\begin{replemma}{lem:strictlySP}
The voting rule $v_{punish}$ is ``strictly strategy-proof'' in the following sense: For every $\alpha > 0$, every $i \in [n]$, every pair of preference orderings $P_i, P_i' \in \P$ with $P_i \neq P_i'$, every $\vec{P}_{-i} \in \P^{n-1}$, and every $\alpha$-coarse utility function $u_i$ that is consistent with $P_i$, we have
\begin{align*}
  \E[u_i(v_{punish}(\vec{P}_{-i},P_i))] \geq \E[u_i(v_{punish}(\vec{P}_{-i},P_i'))] + \Omega(\alpha / n).
\end{align*}
\end{replemma}

The proof of Lemma \ref{lem:strictlySP} \emph{roughly} works as follows. If a voter lies about her preference by swapping two adjacent candidates in her preference ordering, then with probability $1/n$, the voter will be chosen, and she will lose a constant amount of expected utility; this is because the less preferred candidate is now higher and thus will be chosen with higher probability, while the more preferred candidate is now lower and thus will be chosen with lower probability (and the utilities assigned to the two swapped candidates have an $\alpha$ gap between them, since the utility function is $\alpha$-coarse). We show that we can obtain any (false) preference ordering from the true preference ordering by performing a sequence of swaps of adjacent candidates, where the less preferred candidate (according to the true preference) is always swapped upwards; each of these swaps causes the voter to lose $\Omega(\alpha/n)$ expected utility, as described earlier. Thus, the lemma holds. 

\begin{proof}
Let $\alpha > 0$, let $i \in [n]$, let $P_i, P_i' \in \P$ with $P_i \neq P_i'$, let $\vec{P}_{-i} \in \P^{n-1}$, and let $u_i: \C \to [0,1]$ be any $\alpha$-coarse utility function that is consistent with $P_i$. We will show that
\begin{align*}
  \E[u_i(v_{punish}(\vec{P}_{-i},P_i))] \geq \E[u_i(v_{punish}(\vec{P}_{-i},P_i'))] + \Omega(\alpha / n). \tag{1}
\end{align*}
Let $\vec{P} = (\vec{P}_{-i},P_i)$ and $\vec{P'} = (\vec{P}_{-i},P_i')$. Let $a_1, \ldots, a_m$ be the ordering of the candidates in the preference ordering $P_i'$, with $a_1$ being the top (highest-ranked) candidate in $P_i'$. We observe that $\vec{P'}$ can be obtained from $\vec{P}$ by performing the following sequence of swaps of adjacent candidates in voter $i$'s preference ordering (similar to how bubble sort works): We first take the candidate $a_1$ in the preference ordering $P_i$ and move $a_1$ to the top position by repeatedly swapping $a_1$ with the candidate directly above; this makes the top candidate of the resulting preference ordering coincide with the top candidate of $P_i'$. We then take the candidate $a_2$ in the resulting preference ordering and move $a_2$ to the second top position by repeatedly swapping the candidate with the candidate directly above; this makes the top two candidates of the resulting preference ordering coincide with the top two candidates of $P'$. We then take the candidate $a_3$ in the resulting preference ordering and move the candidate to the third top position by repeatedly swapping the candidate with the candidate directly above. It is easy to see that by continuing this process in the natural way, we will eventually get the preference ordering $P_i'$. 

We now analyze how the expected utility $\E[u_i(v_{punish}(\vec{P}_{-i},\cdot))]$ changes as we perform the swaps to get from $P_i$ to $P_i'$ for voter $i$'s preference ordering. We note that for each swap, we are swapping a pair of adjacent candidates, say $x$ and $y$ with $x$ on top of $y$ (before the swap), such that the preference ordering $P_i$ ranks $x$ higher than $y$. Let $Q_i$ and $Q_i'$ denote the two preference orderings for voter $i$ before and after such a swap, respectively. Now, we observe that 
\begin{align*}
& \E[u_i(v_{punish}(\vec{P}_{-i},Q_i'))] - \E[u_i(v_{punish}(\vec{P}_{-i},Q_i))] \\
= \ & \frac{1}{n} [u_i(x) \cdot v_{punish}(x, (\vec{P}_{-i},Q_i')) + u_i(y) \cdot v_{punish}(y, (\vec{P}_{-i},Q_i'))] \\
\ & - \frac{1}{n} [u_i(x) \cdot v_{punish}(x, (\vec{P}_{-i},Q_i)) + u_i(y) \cdot v_{punish}(y, (\vec{P}_{-i},Q_i))] \\
= \ & \frac{1}{n} u_i(x) \cdot [v_{punish}(x, (\vec{P}_{-i},Q_i')) - v_{punish}(x, (\vec{P}_{-i},Q_i))] \\
\ & + \frac{1}{n} u_i(y) \cdot [v_{punish}(y, (\vec{P}_{-i},Q_i')) - v_{punish}(y, (\vec{P}_{-i},Q_i))] \\ 
= \ & \frac{1}{n} u_i(x) \cdot \left(- \frac{1}{\sum_{k=1}^m (m-k)}\right) + \frac{1}{n} u_i(y) \cdot \left(\frac{1}{\sum_{k=1}^m (m-k)}\right) \\
= \ & \frac{1}{n} (u_i(y) - u_i(x)) \cdot \left(\frac{2}{m(m-1)}\right).
\end{align*}
Since the preference ordering $P_i$ ranks $x$ higher than $y$, and since the utility function $u_i$ is consistent with $P_i$, we have $u_i(x) > u_i(y)$, so $u_i(y) - u_i(x) \leq - \alpha$ (since $u_i$ is $\alpha$-coarse). Thus, we have
\begin{align*}
\E[u_i(v_{punish}(\vec{P}_{-i},Q_i'))] - \E[u_i(v_{punish}(\vec{P}_{-i},Q_i))] 
\leq - \Omega(\alpha / n).
\end{align*}
Thus, the expected utility $\E[u_i(v_{punish}(\vec{P}_{-i},\cdot))]$ goes down by at least $\Omega(\alpha / n)$ each time we perform a swap in the sequence of swaps to get from $P_i$ to $P_i'$ for voter $i$'s preference ordering. This implies that $\E[u_i(v_{punish}(\vec{P}_{-i},P_i))] \geq \E[u_i(v_{punish}(\vec{P}_{-i},P_i'))] + \Omega(\alpha / n)$, which shows (1), as required. This completes the proof of the lemma.
\end{proof}

\begin{replemma}{lem:mixIsSP}
Let $v: \P^* \to \Delta(\C)$ be any voting rule that is large-scale $\eps$-strategy-proof w.r.t.~coarse i.i.d.~beliefs, where $\eps(n) = o(1/n^2)$. Let $v_{mix}$ be the voting rule that runs $v$ with probability $1-q(n)$ and runs $v_{punish}$ with probability $q(n)$, where $q(n) = \Omega(n^2 \cdot \eps(n))$. Then, $v_{mix}$ is large-scale strategy-proof w.r.t.~coarse i.i.d.~beliefs.
\end{replemma}

The proof of Lemma \ref{lem:mixIsSP} roughly works as follows. By Lemma \ref{lem:strictlySP}, if a voter lies about her preference, she will gain at most $\eps = o(1/n^2)$ expected utility if $v_{mix}$ runs the voting rule $v$, but she will lose at least $\Omega(\alpha / n)$ expected utility if $v_{mix}$ runs the voting rule $v_{punish}$, where $\alpha$ is the coarseness of her utility function. The probability $q$ that $v_{mix}$ runs $v_{punish}$ is appropriately chosen so that overall, the voter does not gain any expected utility from lying. See \ref{app:punishingVotingRule} for the proof of Lemma \ref{lem:mixIsSP}.

\begin{proof}
Let $p(x)$ be the maximum of $p_{old}(x)$ and $x$, where $p_{old}(\cdot)$ is the $p(\cdot)$ guaranteed by the large-scale $\eps$-strategy-proof w.r.t.~coarse i.i.d.~beliefs property of $v$. Let $\alpha > 0$, let $n \geq p(1/\alpha)$, let $i \in [n]$, let $P_i, P_i' \in \P$ with $P_i \neq P_i'$, let $\phi_i$ be any $\alpha$-coarse i.i.d.~belief, and let $u_i$ be any $\alpha$-coarse utility function that is consistent with $P_i$. Let $\vec{P}_{-i} \sim {\phi_i}^{n-1}$. Since $v$ is large-scale $\eps$-strategy-proof w.r.t.~coarse i.i.d.~beliefs, we have 
\begin{align*}
  \E[u_i(v(\vec{P}_{-i},P_i))] \geq \E[u_i(v(\vec{P}_{-i},P_i'))] - \eps(n).
\end{align*}
By Lemma \ref{lem:strictlySP}, we also have
\begin{align*}
  \E[u_i(v_{punish}(\vec{P}_{-i},P_i))] \geq \E[u_i(v_{punish}(\vec{P}_{-i},P_i'))] + \alpha/n.
\end{align*}
Now, we observe that
\begin{align*}
& \E[u_i(v_{mix}(\vec{P}_{-i},P_i))] \\
= \ & (1-q(n)) \cdot \E[u_i(v(\vec{P}_{-i},P_i))] + q(n) \cdot \E[u_i(v_{punish}(\vec{P}_{-i},P_i))] \\
\geq \ & (1-q(n)) \cdot (\E[u_i(v(\vec{P}_{-i},P_i'))] - \eps(n)) + q(n) \cdot (\E[u_i(v_{punish}(\vec{P}_{-i},P_i'))] + \alpha/n) \\
= \ & (1-q(n)) \cdot \E[u_i(v(\vec{P}_{-i},P_i'))] + q(n) \cdot \E[u_i(v_{punish}(\vec{P}_{-i},P_i'))] - (1-q(n)) \cdot \eps(n) + q(n)\alpha / n \\
= \ & \E[u_i(v_{mix}(\vec{P}_{-i},P_i'))] - (1-q(n)) \cdot \eps(n) + q(n) \alpha / n. \tag{1}
\end{align*}
Now, we observe that by choosing $q(n) = \Omega(n^2 \cdot \eps(n))$ appropriately, we have $-(1-q(n)) \cdot \eps(n) + q(n) \alpha / n \geq -\eps(n) + q(n) / (n^2) \geq 0$ (since $\alpha \geq 1/n$), and the lemma follows.
\end{proof}

\section{Proofs for our Impossibility Result (Theorem \ref{thm:impossibility})}
\label{app:impossibility}

\begin{replemma}{lem:superWeakToStrong}
Let $v: \P^n \to \Delta(\C)$ be any anonymous randomized voting rule that is weakly strategy-proof and satisfies $\eps$-super-weak unanimity. Then, $v$ satisfies $\eps$-strong unanimity.
\end{replemma}

\begin{proof}
Let $x \in \C$, and let $\vec{P} = (P_1, \ldots, P_n) \in \P^n$ such that $top(P_i) = x$ for every $i \in [n]$. We will show that $v(x,\vec{P}) \geq 1 - \eps$. Since $v$ satisfies $\eps$-super-weak unanimity, there exists a preference profile $\vec{P'} = (P'_1, \ldots, P'_n) \in \P^n$ with $top(P'_i) = x$ for every $i \in [n]$, such that $v(x,\vec{P'}) \geq 1-\eps$. Now, we observe that since candidate $x$ is at the top for every voter in both $\vec{P}$ and $\vec{P'}$, we can obtain $\vec{P'}$ from $\vec{P}$ by performing a sequence of swaps of adjacent candidates (in the voters' preference orderings) such that none of the swaps involve candidate $x$. Since $v$ is pairwise responsive (Lemma \ref{lem:basicProperties}), the selection probability of $x$ is not changed by any of the swaps. Thus, we have $v(x,\vec{P}) = v(x,\vec{P'}) \geq 1-\eps$, as required.
\end{proof}

\begin{replemma}{lem:closeToTopsOnly}
Let $v: \P^n \to \Delta(\C)$ be any anonymous randomized voting rule that is weakly strategy-proof and satisfies $\eps$-super-weak unanimity. Then, for every pair of preference profiles $\vec{P}, \vec{P'} \in \P^n$ such that $top(\vec{P}) = top(\vec{P'})$, we have $|v(x,\vec{P}) - v(x,\vec{P'})| \leq m\eps$ for every $x \in \C$.
\end{replemma}

\begin{proof}
Let $\C = \{a_1, \ldots, a_m\}$, let $\vec{P} = (P_1, \ldots, P_n), \vec{P'} = (P'_1, \ldots, P'_n) \in \P^n$ such that $top(\vec{P}) = top(\vec{P'})$, and let $x \in \C$. For $j = 0, \ldots, m$, let $\vec{P}^{(j)} = ({P^{(j)}}_1, \ldots, {P^{(j)}}_n)$ be defined as follows: for each $i \in [n]$, let ${P^{(j)}}_i = P'_i$ if $top(P_i) \in \{a_1, \ldots, a_j\}$, and let ${P^{(j)}}_i = P_i$ otherwise. We note that $\vec{P}^{(0)} = \vec{P}$ and $\vec{P}^{(m)} = \vec{P'}$. We will show that for every $0 \leq j \leq m-1$, we have $|v(x,\vec{P}^{(j)}) - v(x,\vec{P}^{(j+1)})| \leq \eps$. The lemma then immediately follows by repeated application of the triangle inequality.

Let $0 \leq j \leq m-1$. From the definition of $\vec{P}^{(j)}$ and $\vec{P}^{(j+1)}$, we see that $\vec{P}^{(j)}$ and $\vec{P}^{(j+1)}$ only differ in the components $i \in [n]$ for which $top(P_i) = a_{j+1}$; let $I$ be the set of all such $i \in [n]$. For each $i \in I$, we have ${\vec{P}^{(j)}}_i = P_i$ and ${\vec{P}^{(j+1)}}_i = P'_i$. However, since $top(\vec{P}) = top(\vec{P'})$, $top(P_i) = a_{j+1}$ is at the top of both ${\vec{P}^{(j)}}_i$ and ${\vec{P}^{(j+1)}}_i$ for every $i \in I$, so $\vec{P}^{(j+1)}$ can be obtained from $\vec{P}^{(j)}$ via a sequence of swaps of adjacent candidates (in the preference orderings of voters in $I$) that do not involve candidate $a_{j+1}$. Let $\vec{Q}^{(0)}, \ldots, \vec{Q}^{(r)}$ be any sequence of preference profiles generated by such a sequence of swaps, where $\vec{Q}^{(0)} = \vec{P}^{(j)}$ and $\vec{Q}^{(r)} = \vec{P}^{(j+1)}$. Now, we observe that
\begin{align*}
v(x,\vec{P}^{(j+1)}) - v(x,\vec{P}^{(j)}) = \sum_{k=0}^{r-1} \left( v(x,\vec{Q}^{(k+1)}) - v(x,\vec{Q}^{(k)}) \right). \tag{1}
\end{align*}
For each $0 \leq k \leq r$, let $\vec{Q'}^{(k)}$ be the same as $\vec{Q}^{(k)}$ except that for every $i \in [n] \setminus I$, the candidate $a_{j+1}$ in the preference ordering ${\vec{Q}^{(k)}}_i$ is moved to the top of the preference ordering; we note that $a_{j+1}$ is at the top of all the preference orderings in $\vec{Q'}^{(k)}$. We will now show that for each $0 \leq k \leq r-1$, we have
\begin{align*}
v(x,\vec{Q}^{(k+1)}) - v(x,\vec{Q}^{(k)}) = v(x,\vec{Q'}^{(k+1)}) - v(x,\vec{Q'}^{(k)}). \tag{2}
\end{align*}
Let $0 \leq k \leq r-1$. Recall that $\vec{Q}^{(k+1)}$ is obtained from $\vec{Q}^{(k)}$ by performing a swap of adjacent candidates in the preference ${\vec{Q}^{(k)}}_i$ for some $i \in I$, and the swap does not involve candidate $a_{j+1}$. If the swap does not involve the candidate $x$, then since $v$ is pairwise responsive (Lemma \ref{lem:basicProperties}), both sides of (2) are 0, which proves (2). Thus, we now assume that the swap involves the candidate $x$. If the swap moves $x$ upwards, then since $v$ is pairwise isolated (Lemma \ref{lem:basicProperties}), (2) holds. If the swap moves $x$ downwards, then the swap moves some other candidate upwards, say, candidate $y$. Then, since $v$ is pairwise responsive and the selection probabilities of the candidates must add up to 1, we have $v(x,\vec{Q}^{(k+1)}) - v(x,\vec{Q}^{(k)}) = -( v(y,\vec{Q}^{(k+1)}) - v(y,\vec{Q}^{(k)}) )$. Since $v$ is pairwise isolated, we have $v(y,\vec{Q}^{(k+1)}) - v(y,\vec{Q}^{(k)}) = v(y,\vec{Q'}^{(k+1)}) - v(y,\vec{Q'}^{(k)}) = - (v(x,\vec{Q'}^{(k+1)}) - v(x,\vec{Q'}^{(k)}))$, where the last equality again uses the fact that $v$ is pairwise responsive. Combining the above equations yields (2), as required. Thus, we have shown (2).

Now, combining (1) and (2), we get
\begin{align*}
v(x,\vec{P}^{(j+1)}) - v(x,\vec{P}^{(j)}) = \sum_{k=0}^{r-1} \left( v(x,\vec{Q'}^{(k+1)}) - v(x,\vec{Q'}^{(k)}) \right) = v(x,\vec{Q'}^{(r)}) - v(x,\vec{Q'}^{(0)}).
\end{align*}
To complete the proof of the lemma, it suffices to show that $|v(x,\vec{Q'}^{(r)}) - v(x,\vec{Q'}^{(0)})| \leq \eps$. 

Since $v$ satisfies $\eps$-super-weak unanimity, we have that by Lemma \ref{lem:superWeakToStrong}, $v$ also satisfies $\eps$-strong unanimity. Thus, we have $v(a_{j+1},\vec{Q'}^{(r)}) \geq 1-\eps$ and $v(a_{j+1},\vec{Q'}^{(0)}) \geq 1-\eps$, since candidate $a_{j+1}$ is at the top of all the preference orderings in $\vec{Q'}^{(r)}$ and $\vec{Q'}^{(0)}$.

If $x = a_{j+1}$, then $|v(x,\vec{Q'}^{(r)}) - v(x,\vec{Q'}^{(0)})| = |v(a_{j+1},\vec{Q'}^{(r)}) - v(a_{j+1},\vec{Q'}^{(0)})| \leq \eps$, since $v(a_{j+1},\vec{Q'}^{(r)}) \in [1-\eps,1]$ and $v(a_{j+1},\vec{Q'}^{(0)}) \in [1-\eps,1]$. 

If $x \neq a_{j+1}$, then $v(x,\vec{Q'}^{(r)}) \in [0,\eps]$ and $v(x,\vec{Q'}^{(0)}) \in [0,\eps]$ (since $v(a_{j+1},\vec{Q'}^{(r)}) \geq 1-\eps$ and $v(a_{j+1},\vec{Q'}^{(0)}) \geq 1-\eps$), so $|v(x,\vec{Q'}^{(r)}) - v(x,\vec{Q'}^{(0)})| \leq \eps$, as required. 

This completes the proof of the lemma.
\end{proof}

\begin{replemma}{lem:timesAtTop}
Let $v: \P^n \to \Delta(\C)$ be any anonymous randomized voting rule that is weakly strategy-proof and satisfies $\eps$-super-weak unanimity. Then, for every candidate $x \in \C$, and every pair of preference profiles $\vec{P} = (P_1, \ldots, P_n), \vec{P'} = (P'_1, \ldots, P'_n) \in \P^n$ such that $|\{i \in [n] : top(P_i) = x\}| = |\{i \in [n] : top(P'_i) = x\}|$, we have $|v(x,\vec{P}) - v(x,\vec{P'})| \leq 2m\eps$.
\end{replemma}

\begin{proof}
Let $x \in \C$, and let $\vec{P} = (P_1, \ldots, P_n), \vec{P'} = (P'_1, \ldots, P'_n) \in \P^n$ such that $|\{i \in [n] : top(P_i) = x\}| = |\{i \in [n] : top(P'_i) = x\}|$. Let $k = |\{i \in [n] : top(P_i) = x\}|$. Since $v$ is anonymous, we can assume without loss of generality that for $i = 1, \ldots, k$, we have $top(P_i) = x$ and $top(P'_i) = x$. Then, for $i = k+1, \ldots, n$, we have $top(P_i) \neq x$ and $top(P'_i) \neq x$. Let $a_1, \ldots, a_m$ be any ordering of the set of candidates $\C$ such that $a_m = x$, where $a_1$ is the highest-ranked candidate and $a_m$ is the lowest-ranked candidate. Let $\vec{P^*} = (P^*_1, \ldots, P^*_n)$ be the preference profile defined as follows: for $i = 1, \ldots, k$, let $top(P^*_i) = x$ and the rest of $P^*_i$ is ordered according to the ordering $a_1, \ldots, a_m$; for $i = k+1, \ldots, n$, let $P^*_i$ be the ordering $a_1, \ldots, a_m$. 

We first show that $|v(x,\vec{P}) - v(x,\vec{P^*})| \leq m\eps$. We will perform a sequence of operations on $\vec{P}$ to obtain $\vec{P^*}$, and we will analyze how these operations affect the selection probability $v(x,\cdot)$ of $x$. We start with $\vec{P}$, and for every $i \in \{k+1, \ldots, n\}$, we simultaneously move the candidate $x$ in $P_i$ to the bottom. Since $top(P_i) \neq x$ for $i = k+1, \ldots, n$, by Lemma \ref{lem:closeToTopsOnly}, this operation changes the selection probability of $x$ by at most $m\eps$. Now, we observe that from this new preference profile, we can obtain $\vec{P^*}$ by performing a sequence of swaps of adjacent candidates in the preference orderings such that none of the swaps involve candidate $x$. Since $v$ is pairwise responsive (Lemma \ref{lem:basicProperties}), none of these swaps change the selection probability of $x$. Thus, the overall change in the selection probability of $x$ is at most $m\eps$, so $|v(x,\vec{P}) - v(x,\vec{P^*})| \leq m\eps$, as required.

By a similar argument, we also have $|v(x,\vec{P'}) - v(x,\vec{P^*})| \leq m\eps$. Thus, by the triangle inequality, we have $|v(x,\vec{P}) - v(x,\vec{P'})| \leq 2m\eps$. This completes the proof of the lemma.
\end{proof}

\begin{repclaim}{claim:candidatesAreAnonymous}
For every pair of candidates $x,y \in \C$ and every $j \in \{0, \ldots, n\}$, we have $|v'(x,j) - v'(y,j)| \leq 14m\eps$.
\end{repclaim}

\begin{proof}
Let $x,y \in \C$ and $j \in [n]$. Let $z$ be any candidate in $\C \setminus \{x,y\}$. Let $\vec{P} = (P_1, \ldots, P_n)$ be any preference profile such that for every $i \in [n]$, we have $top(P_i) = z$ and candidate $x$ is directly below $z$. Now, for $i = 1, \ldots, j$, we swap the candidates $z$ and $x$ in $P_i$ so that $x$ is now at the top; let's call the resulting preference profile $\vec{P'}$. Since the swaps we performed only involved the candidates $z$ and $x$, and since $v$ is pairwise responsive, we have $v(x,\vec{P'}) - v(x,\vec{P}) = - (v(z,\vec{P'}) - v(z,\vec{P})) = v'(z,n) - v'(z,n-j) + \delta$, where $|\delta| \leq 4m\eps$ (by Claim \ref{claim:vCloseTov'}). Now, we note that $|v(x,\vec{P})| \leq \eps$ (by strong unanimity of $v$), so $v(x,\vec{P'}) = v'(z,n) - v'(z,n-j) + \delta'$, where $|\delta'| \leq 5m\eps$. Thus, by Claim \ref{claim:vCloseTov'}, we have $v'(x,j) = v'(z,n) - v'(z,n-j) + \gamma$, where $|\gamma| \leq 7m\eps$.

By a similar argument but where we use $y$ instead of $x$, we also have $v'(y,j) = v'(z,n) - v'(z,n-j) + \gamma'$, where $|\gamma'| \leq 7m\eps$. Thus, we have
\begin{align*}
|v'(x,j) - v'(y,j)| = |\gamma - \gamma'| \leq 14m\eps.
\end{align*}
This completes the proof of the claim.
\end{proof}

\begin{repclaim}{claim:slidingWindow}
Let $x \in \C$, let $j,j' \in \{0, \ldots, n-1\}$, and let $\ell \in [n]$ such that $j + \ell \leq n$ and $j' + \ell \leq n$. Then,
\begin{align*}
v'(x,j+\ell) - v'(x,j) = v'(x,j'+\ell) - v'(x,j') + \delta
\end{align*}
for some $\delta \in \R$ such that $|\delta| \leq 64m\eps$.
\end{repclaim}

\begin{proof}
We first show that $v'(x,j+\ell) - v'(x,j) = v'(x,\ell) - v'(x,0) + \gamma$, where $|\gamma| \leq 32m\eps$. Let $y,z \in \C$ such that $x,y,z$ are all distinct. Let $\vec{P} = (P_1, \ldots, P_n)$ be any preference profile such that for $i = 1, \ldots, j$, we have $top(P_i) = x$, and for $i = j+1, \ldots, j+\ell$, we have $top(P_i) = y$ and candidate $x$ is directly below $y$, and for $i = j+\ell+1, \ldots, n$, we have $top(P_i) = z$. Now, for $i = j+1, \ldots, j+\ell$, we swap the candidates $y$ and $x$ in $P_i$ so that $x$ is now at the top; let's call the resulting preference profile $\vec{P'}$. Since the swaps we performed only involved the candidates $y$ and $x$, and since $v$ is pairwise responsive, we have $v(x,\vec{P'}) - v(x,\vec{P}) = - (v(y,\vec{P'}) - v(y,\vec{P})) = v'(y,\ell) - v'(y,0) + \delta'$, where $|\delta'| \leq 4m\eps$. Now, by Claim \ref{claim:candidatesAreAnonymous}, we have $|v'(y,\ell) - v'(x,\ell)| \leq 14m\eps$ and $|v'(y,0) - v'(x,0)| \leq 14m\eps$, so $v(x,\vec{P'}) - v(x,\vec{P}) = v'(x,\ell) - v'(x,0) + \gamma$, where $|\gamma| \leq 32m\eps$. 

By a similar argument, we also have $v'(x,j'+\ell) - v'(x,j') = v'(x,\ell) - v'(x,0) + \gamma'$, where $|\gamma'| \leq 32m\eps$. The claim then follows by the triangle inequality.
\end{proof}

\begin{repclaim}{claim:closeToRandomDictatorship}
Let $x \in \C$. For every $j \in \{0, \ldots, n\}$, we have $|v'(x,j) - \frac{j}{n}| \leq O(\eps)$. 
\end{repclaim}

\begin{proof}
Since $v$ satisfies $\eps$-strong unanimity, we have $v'(x,n) \geq 1 - \eps$ and $v'(x,0) \leq \eps$. Now, for every $q \leq n/2$, we have $v'(x,2q) - v'(x,q) = v'(x,q) - v'(x,0) + O(\eps)$ by Claim \ref{claim:slidingWindow}, so 
\begin{align*}
v'(x,2q) = 2v'(x,q) + O(\eps). \tag{1}
\end{align*}
Let $q \in \{0, \ldots, n\}$ such that $r_q := |v'(x,q)-q/n|$ is maximal. We will show that $r_q \leq O(\eps)$.

Case 1: $q \leq n/2$. By (1), we have
\begin{align*}
r_{2q} = |v'(x,2q) - 2q/n| = |2v'(x,q)+O(\eps) - 2q/n| = 2r_q - O(\eps).
\end{align*}
By the maximality of $r_q$, we have $r_{2q} \leq r_q$, so $2r_q - O(\eps) \leq r_q$, so $r_q \leq O(\eps)$, as required.

Case 2: $q > n/2$. Let $q' = n - q$. Then, by Claim \ref{claim:slidingWindow}, we have $v'(x,q')-v'(x,0) = v'(x,n) - v'(x,q) + O(\eps)$, so $v'(x,q') = 1 - v'(x,q) + O(\eps)$. Then, we have 
\begin{align*}
r_{q'} = |v'(x,q')-q'/n| = |1 - v'(x,q) + O(\eps) - (n-q)/n| = r_{q} - O(\eps). 
\end{align*}
Now, by (1), we have
\begin{align*}
r_{2q'} = |v'(x,2q') - 2q'/n| = |2v'(x,q') + O(\eps) - 2q'/n| = 2r_{q'} - O(\eps) = 2r_q - O(\eps).
\end{align*}
By the maximality of $r_q$, we have $r_{2q'} \leq r_q$, so $2r_q - O(\eps) \leq r_q$, so $r_q \leq O(\eps)$, as required.
\end{proof}

\end{document}